# DesignCon 2016

## 56 Gbps PCB Design Strategies for Clean, Low-Skew Channels

Michael J. Degerstrom, Mayo Clinic
degerstrom.michael@mayo.edu

Chad M. Smutzer, Mayo Clinic
smutzer.chad@mayo.edu

Barry K. Gilbert, Mayo Clinic
gilbert.barry@mayo.edu, 507-284-4056

Erik S. Daniel, Mayo Clinic

## Abstract

Although next generation (>28 Gbps) SerDes standards have been contemplated for several years, it has not been clear whether PCB structures supporting 56 Gbps NRZ will be feasible and practical. In this paper, we assess a number of specific PCB design strategies (related to pin-field breakouts, via stubs, and fiber weave skew) both through simulation and through measurement of a wide range of structures on a PCB test vehicle. We demonstrate that conventional approaches in many cases will not be sufficient, but that modest (manufacturable) design changes can enable low-skew 56 Gbps NRZ channels having acceptable insertion and return loss.

## Biographies

**Michael J. Degerstrom** received a BSEE from the University of Minnesota. Mike is currently a Senior Engineer at the Mayo Clinic Special Purpose Processor Development Group. His primary area of research and design has been in the specialty of signal and power integrity.

**Chad M. Smutzer** received a BSEE from the University of Iowa in Iowa City. He is currently a Senior Engineer at the Mayo Clinic Special Purpose Processor Development Group where he performs signal and power integrity analysis.

**Barry K. Gilbert** received a BSEE from Purdue University (West Lafayette, IN) and a Ph.D. in physiology and biophysics from the University of Minnesota (Minneapolis, MN). He is currently Director of the Special Purpose Processor Development Group, directing research efforts in high performance electronics and related areas.

**Erik S. Daniel** received a BA in physics and mathematics from Rice University (Houston, TX) in 1992 and a Ph.D. degree in solid state physics from the California Institute of Technology (Pasadena, CA) in 1997. He contributed to a wide array of research and development efforts at Mayo Clinic from 1998-2015, most recently serving as the Deputy Director of the Special Purpose Processor Development Group. Dr. Daniel is currently a member of the Senior Scientist staff at HRL Laboratories, continuing to pursue research and development of high performance electronic systems and components.

## Introduction

The demand for higher data rate digital communication is expected to continue. Trends show, for example, based on emerging Ethernet standards, an approximate doubling of digital throughput every two years [1, p33]. This data rate increase is typically met both with a larger number of links for a particular standard and also with a significant speed increase per link.

Presently, 28 gigabit per second (Gb/s) non-return-to-zero (NRZ) data rates are emerging. As described in this manuscript, we are working to understand the printed circuit board (PCB) design challenges for twice that data rate. We are specifically targeting 56 Gb/s with NRZ coding on a single lane – not to be confused, for example, with four lanes of 14 Gb/s. Nor will we discuss 28 gigabaud per second (GBaud/s) using more advanced transmitter modulation and receiver circuitry such as 4-level pulse-amplitude-modulation (PAM-4) that offers the same throughput as 56 Gb/s NRZ.

We have designed a test board to evaluate various capabilities, although this paper will concentrate on the capability for supporting direct, otherwise unmodulated 56 Gb/s NRZ signaling rates. Our typical boards require provision for high current flows to support the power up of low-voltage cores. Therefore antipads, i.e., the metal plane voids surrounding a signal via and pad, cannot be too large, since large openings in the planes would cause substantial DC voltage drops within the pin-fields supporting high-current processors. Conversely, antipads will need to be as large as possible to limit pin-field capacitances for 56 Gb/s data rates.

One result of our study indicates that, for 56 Gb/s applications, PCB structure analysis is anticipated to be very dependent on statistical manufacturing variations that will allow us to quantify board-level performance.

## Discussion

To facilitate readability in the following material we now introduce a special term, "$56G_{NRZ}$" to explicitly mean a data rate of 56 Gb/s with NRZ coding. In addition, we will refer to differing PCB layers within a test board, as follows: specific layers will be called out simply as "LN", e.g., "L7" refers to the seventh PCB layer. Note that by convention PCB layer numbering starts at the top of the board and increases sequentially to the bottom layer. Five boards were measured in our laboratory and categorized as SN1 through SN5.

At present there is significant concentration on 28 GBaud/s PAM-4 in order to meet the 56 Gb/s throughput per lane requirement. Some of the results discussed below can be applied to 28 GBaud/s PAM-4 applications. However, for $56G_{NRZ}$ we will describe the need for PCBs having exceptional capabilities, whereas for 28 Gb/s PAM-4 it is widely accepted that a less-capable PCB can be utilized. When we began the design of a test board, $56G_{NRZ}$ was being studied by serial/de-serializer (SerDes) circuit designers for backplane channel applications [2]. In addition, 56 Gb/s signaling using duo-binary encoding was also being studied [3]. It is our intent that this manuscript will provide design insights regardless of the targeted data rates or coding schemes.

We begin with a detailed description of our test board. Next we will discuss the performance for our baseline "control" channels. We then review various design techniques to ascertain if performance improvements can be obtained, and to describe situations in which performance is lost. A study of impedance control is presented. Next we discuss PCB layer shifting behavior and the effect that this layer shifting has on electrical performance. We conclude with a discussion on PCB channel skew.

## Test Board Description

We designed and had built a test board to demonstrate a proof-of-concept implementation for actual next-generation system boards. The goals of this test board were as follows: 1) determine the ease of manufacturability, 2) demonstrate three or more signal layers supporting $56G_{NRZ}$, 3) identify four or more additional signal layers supporting up to 16 Gb/s NRZ data-rates, 5) incorporate roughly 16 power/ground layers to supply many voltage domains, including the ability to support multiple processors having peak core currents up to 80A, and 6) verify the option to fit within a Peripheral Component Interconnect Express (PCIe) form factor.

Working with our board fabricator, who advised us on available and compatible material sets and overall ease of manufacturability, we devised the stack-up presented in Figure 1. The primary constraint of conforming to a PCIe form factor was to maintain the card edge connector at the specified 0.062" thickness. It was obvious, with our requirements for the multiple power, ground, and signal layers that we had to employ a design such as a "stepped-edge" approach to meet the edge connector thickness constraint. Therefore, we elected to have the board manufactured as two separate boards which were then interconnected with plated-through-hole (PTH) vias. The lower sub-lamination was designed to have the targeted 0.062" thickness. There were no thickness constraints on the top sub-lamination other than the goal of keeping the overall board thin enough such that the PTH via diameters could be constrained to 10 mil diameters and still meet the fabricator's board thickness to via diameter aspect ratio guidance. Based on initial electromagnetic modeling, we believed that using via diameters larger than 10 mils would make it difficult to achieve our $56G_{NRZ}$ performance goals.

Since it is difficult to reach lower layers with laser vias, for the PCB layers immediately below L3 we implemented three buried capacitance (BC) layer pairs, for a total of six 1-ounce copper layers, to provide good DC and reasonable AC power delivery. Many high-speed board designers elect to use back-drilling to remove via stubs to achieve high performance vias. Often the quoted remaining back-drill stub length is 10-15 mils due to uncertainties of layer depths and other considerations. The electrical performance of the residual back-drilled stub of 15 mils would have to be investigated to determine viability for $56G_{NRZ}$.

Each of our sub-laminates used PTH vias that became mechanically drilled blind vias after laminating the two sub-laminations into the final PCB. With our thickest sub-lamination at 0.062", our selection of an 8 mil mechanical via drill was easily implemented, since the via aspect ratio was approximately 8:1, which allowed for very easy via manufacturing. Our top sub-lamination, layers 1-12, had good candidate high performance vias at L10 and at L12 with a 10 mil stub and no stub, respectively. These two vias, along with a L3 laser-blind via, yielded a total of three high-performance routing layers in the top sub-laminate.

We achieved a fourth via with a short stub at layer 24 of the board. This via requires a 10 mil drill and is roughly twice as long as the mechanically-drilled vias in the top sub-lamination. Three additional stripline layers in the lower sub-lamination have longer via stubs that are adequate for slower speed serializer/deserializer (SerDes) and other needs. There are two additional BC layer pairs in the lower sub-laminate to assist in power delivery for the myriad voltage supplies that are required by many SerDes-capable modern processors. We also employ laser-blind vias from the bottom of the top lamination (L12 to L10) and on the top of the lower lamination (L13 to L15) and from the bottom of the lower lamination (L26 to L24) to reach nearby stripline layers. A mechanical 8 mil drill is used in the bottom lamination as well. Since all of the high-speed signals originate from top-mounted devices, these additional vias generally have less utility. However, some of the interfacing to the PCIe edge connector can be implemented in the lower laminate. As many of the PCIe channels require DC-blocking capacitors, it is useful to have the laser-blind and mechanical blind vias on the lower laminate, since assembly of smaller back-side components is readily accomplished. Below, we will also describe "tricks" to utilize vias in order to reduce or eliminate via stubs.

Selecting the materials for a dense, high layer-count, $56G_{NRZ}$–capable PCB is not a trivial task. Clearly, low-loss dielectric materials are required for the high-speed signaling. However the complete material systems must also enable very high bandwidth vias. Signal vias in a 1 mm pin-field are typically overly-capacitive in a practical PCB. Therefore we desire materials with the lowest possible dielectric constant and with excellent mechanical stability to minimize the PCB layer shifting that effectively moves antipads closer to the vias, in turn increasing capacitance. Dielectric constants are required to be low so that thin cores and pre-pregs can be used to implement a stripline with an impedance of 50 ohms. The thin dielectric layers provide a thinner board, allowing for smaller mechanically-drilled vias which reduce via capacitance. Conversely, the need for lower loss striplines drives the requirement for wider striplines to minimize high-frequency copper losses. For this test board we chose to use differential pair striplines approximately 5 mils wide with a pitch of 12 mils.

We selected Isola's Tachyon-100G (Tachyon) as the targeted core and pre-preg system. For the BC layers we followed the board manufacturer's recommendation to use Isola's Ultra EC25 cores, which are reinforced materials that were thought to reduce layer shifting. These layers had about 1 mil of dielectric thickness with a dielectric constant of approximately 4.0 that should minimally capacitively load the high performance vias that are drilled through this BC material.

| Lamination | | | | | | | | | |
|---|---|---|---|---|---|---|---|---|---|
| Layer # | Layer Usage | Dielectric | Copper Weight (oz) | Copper Thickness (mil) | Copper Thickness (mm) | Dielectric Thickness (mil) | Dielectric Thickness (mm) | Dk @ 1 GHz | Layer # |
| | Soldermask | Green | | | | 1 | 0.02 5 | 3.5 | |
| 1 | Traces | | 0.25 * | 2.3 0 | 0.05 8 | | | | 1 |
| | Prepreg | 1078 - Isola Tachyon 100G | | | | 3.34 | 0.085 | 3 | |
| 2 | Plane | | 0.5 | 0.60 | 0.015 | | | | 2 |
| | Core | Lam Tachyon100G.004 2x1035 5/5VLP2 24x18 | | | | 4 | 0.102 | 3 | |
| 3 | Traces | | 0.5 | 0.60 | 0.015 | | | | 3 |
| | Prepreg | 1067 / 1067 - Isola Tachyon 100G | | | | 5.24 | 0.133 | 3 | |
| 4 | Plane | | 1 | 1.20 | 0.030 | | | | 4 |
| | Core | Lam, Isola Ultra EC25 1/1 Cu 18x24 | | | | 1 | 0.025 | 4 | |
| 5 | Plane | | 1 | 1.20 | 0.030 | | | | 5 |
| | Prepreg | 1035 / 1035 - Isola Tachyon 100G | | | | 4.76 | 0.121 | 3 | |
| 6 | Plane | | 1 | 1.20 | 0.030 | | | | 6 |
| | Core | Lam, Isola Ultra EC25 1/1 Cu 18x24 | | | | 1 | 0.025 | 4 | |
| 7 | Plane | | 1 | 1.20 | 0.030 | | | | 7 |
| | Prepreg | 1035 / 1035 - Isola Tachyon 100G | | | | 4.76 | 0.121 | 3 | |
| 8 | Plane | | 1 | 1.20 | 0.030 | | | | 8 |
| | Core | Lam, Isola Ultra EC25 1/1 Cu 18x24 | | | | 1 | 0.025 | 4 | |
| 9 | Plane | | 1 | 1.20 | 0.030 | | | | 9 |
| | Prepreg | 1067 / 1067 - Isola Tachyon 100G | | | | 5.24 | 0.133 | 3 | |
| 10 | Traces | | 0.5 | 0.6 0 | 0.01 5 | | | | 10 |
| | Core | Lam Tachyon100G.004 2x1035 5/5VLP2 24x18 | | | | 4 | 0.102 | 3 | |
| 11 | Plane | | 0.5 | 0.60 | 0.015 | | | | 11 |
| | Prepre | 1078 - Isola Tachyon 100G | | | | 4.04 | 0.103 | 3 | |
| 12 | Traces | | 0.25* | 1.40 | 0.036 | | | | 12 |
| | Prepreg | 1078 / 1078 - Isola Tachyon 100G | | | | 7.06 | 0.179 | 3 | |
| 13 | Plan e | | 0.25 * | 1.4 0 | 0.03 6 | | | | 13 |
| | Prepreg | 1078 - Isola Tachyon 100G | | | | 4.04 | 0.103 | 3 | |
| 14 | Plane | | 0.5 | 0.60 | 0.015 | | | | 14 |
| | Core | Lam Tachyon100G.004 2x1035 5/5VLP2 24x18 | | | | 4 | 0.102 | 3 | |
| 15 | Traces | | 0.5 | 0.60 | 0.015 | | | | 15 |
| | Prepreg | 1035 / 1067 - Isola Tachyon 100G | | | | 5 | 0.127 | 3 | |
| 16 | Plane | | 0.5 | 0.60 | 0.015 | | | | 16 |
| | Core | Lam Tachyon100G.004 2x1035 5/5VLP2 24x18 | | | | 4 | 0.102 | 3 | |
| 17 | Traces | | 0.5 | 0.60 | 0.015 | | | | 17 |
| | Prepreg | 1035 / 1067 - Isola Tachyon 100G | | | | 4.94 | 0.125 | 3 | |
| 18 | Plane | | 1 | 1.20 | 0.030 | | | | 18 |
| | Core | Lam, Isola Ultra EC25 1/1 Cu 18x24 | | | | 1 | 0.025 | 4 | |
| 19 | Plane | | 1 | 1.20 | 0.030 | | | | 19 |
| | Prepreg | 1080 - Isola Tachyon 100G | | | | 3.36 | 0.085 | 3 | |
| 20 | Plan e | | 1 | 1.2 0 | 0.03 0 | | | | 20 |
| | Core | Lam, Isola Ultra EC25 1/1 Cu 18x24 | | | | 1 | 0.025 | 4 | |
| 21 | Plane | | 1 | 1.20 | 0.030 | | | | 21 |
| | Prepreg | 1067 / 1035 - Isola Tachyon 100G | | | | 4.94 | 0.125 | 3 | |
| 22 | Traces | | 0.5 | 0.60 | 0.015 | | | | 22 |
| | Core | Lam Tachyon100G.004 2x1035 5/5VLP2 24x18 | | | | 4 | 0.102 | 3 | |
| 23 | Plane | | 0.5 | 0.60 | 0.015 | | | | 23 |
| | Prepreg | 1067 / 1035 - Isola Tachyon 100G | | | | 5 | 0.127 | 3 | |
| 24 | Traces | | 0.5 | 0.6 0 | 0.01 5 | | | | 24 |
| | Core | Lam Tachyon100G.004 2x1035 5/5VLP2 24x18 | | | | 4 | 0.102 | 3 | |
| 25 | Plane | | 0.5 | 0.6 0 | 0.01 5 | | | | 25 |
| | Prepreg | 1078 - Isola Tachyon 100G | | | | 3.34 | 0.085 | 3 | |
| 26 | Traces | | 0.25* | 2.30 | 0.058 | | | | 26 |
| | Soldermask | Green | | | | 1 | 0.02 5 | 3.5 | |
| | | Total Board Thickness | | 122.6 6 | mils | | | | |

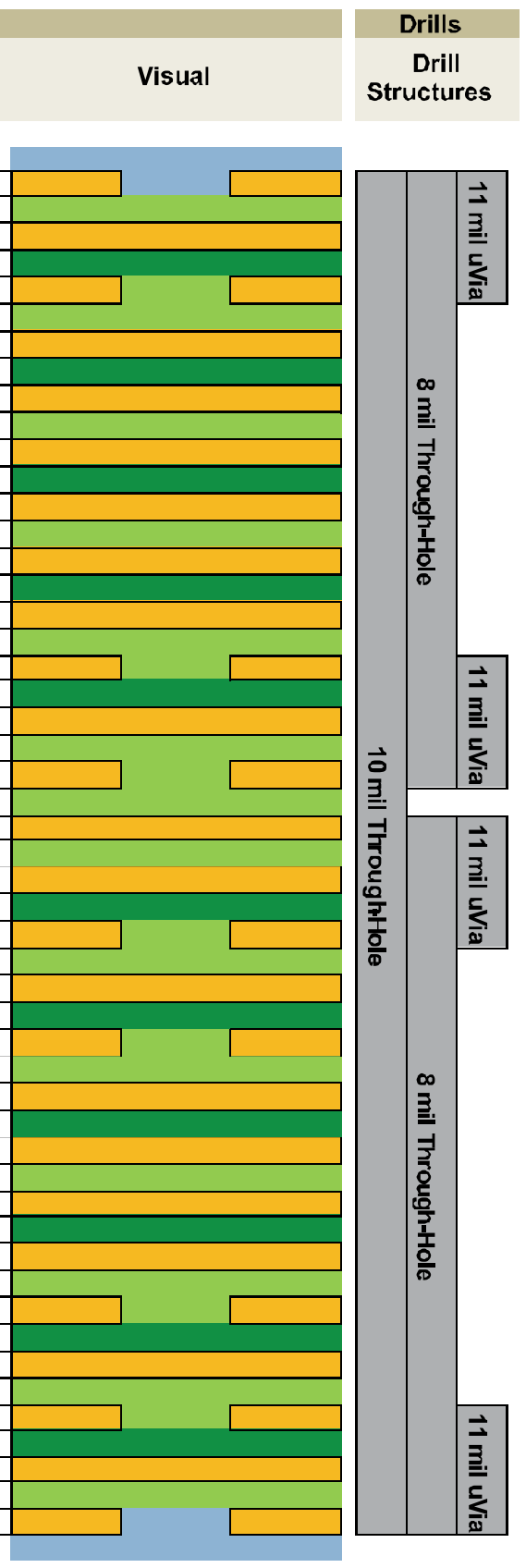

**Figure 1 - Test board stack-up definition [45072 ]**

The top view of the test board appears in Figure 2. The left side of the board has many 4.2" stripline and microstrip test structures running both horizontally and vertically. Along the upper middle of the board are 0.2" "thru" test structures that could be used to de-embed the 4.2" structures. However, the primary intended use for the 4.2" structures was to emulate the PCB-portion of an intra-board SerDes channel so that measurements could validate conformance to $56G_{NRZ}$ electrical performance requirements.

The artwork for three copies of the board are stepped and repeated ("3-up") for imaging on a standard 18" by 24" panel. Each board artwork is rotated 10 degrees from orthogonal to obtain structures of improved uniformity (compared to the non-rotated case) and to mitigate fiber-weave-skew (FWS) effects for differential structures.

In the middle of the test board are two sets of 21 differential pairs to test for FWS. The artwork for the upper set of these structures was rotated 10 degrees from orthogonal and was rotated back during artwork placement of each board on the panel, so that this upper set represents the non-rotated FWS test patterns. The lower set of 21 differential pairs has no rotation and will be the test structures that have 10 degree rotation when the artwork is rotated and placed on the panel.

We also incorporated layer alignment structures both in the Southwest and Northeast corners of this board. A variety of other structures were incorporated into this board that will not be described in this manuscript.

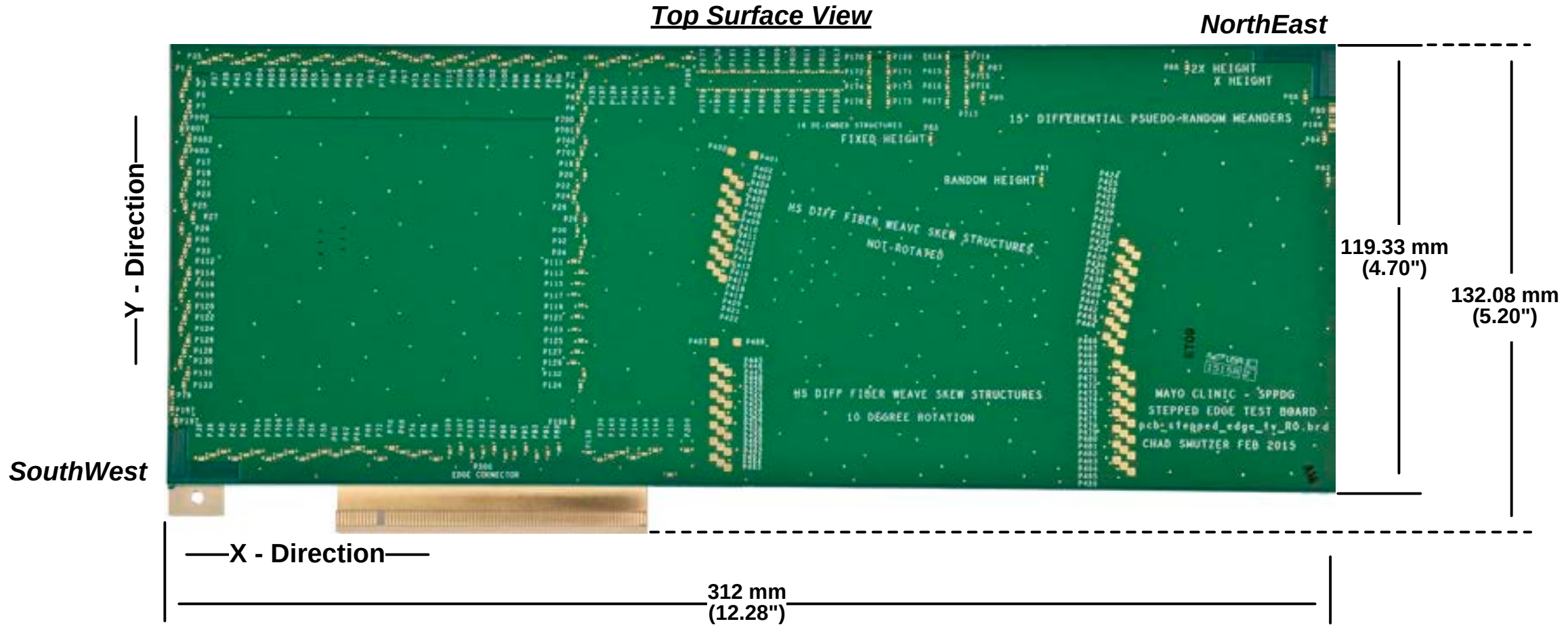

**Figure 2 – Top and view of fabricated test board [44974 ]**

We employ high-frequency microwave probes rather than high-performance connectors to measure the test structures. Probes gave us the best ability to emulate the devices under test (DUTs) as actual SerDes intra-board channels, though neglecting the BGA portions of these channels. Specifically, since we calibrate the vector network analyzer (VNA) to the probe tips, what we measure is the PCB channel performance. The probe interface design is documented in Figure 3. We identified a potentially satisfactory probe launch, illustrated in Figure 3-a [4]. This approach utilized a flooded ground outer layer and dual G-S-G probes straddling the signal pads. The probe pad design had to be compatible with a flooded ground plane on the top surface, as all non-plane layers are flooded with metal on our board to achieve better layer uniformity. However, flexibility of the top-layer antipad shapes was required to match the antipad designs that would be employed in an actual system board. Generally the outer layer antipads are made larger than those used for inner layers to reduce fringing capacitance between signal pads and the flooded top layer – a problem not generally required for inner layers, since all non-functional pads are removed along the signal via barrel. We employed a short microstrip breakout into the pin-field as shown in Figure 3-b. We simulated this structure with the model shown in Figure 3-c. Figure 3-d illustrates the simulated return loss for this model of better than -12 dB at the Nyquist frequency of 28 GHz. The return loss performance was determined using the broadband return loss (BRL) approach [5]. We did not review return losses at a particular frequency and instead, throughout this paper, imply BRL performance when describing return loss results. Note that we did not attempt to achieve the absolute best electrical performance of the probed launches, but rather to emulate the actual pin-field performance, with the constraint that electrical contact must be provided so that the pin-field could be measured without changing the performance as it would in an actual system board.

GGB G-S-G 250 DUAL microwave probes were used to mate with the 250 micron pitch G-S-G-G-S-G probe pattern. These probes were rated to 40 GHz, and therefore we did not take measurements above this frequency. We recognize that measurements to higher frequencies for $56G_{NRZ}$ applications may be important to detect deleterious electrical behavior above 40 GHz, and thus identifying higher bandwidth probes to measure PCB channel performance is a future investigation that we intend to undertake.

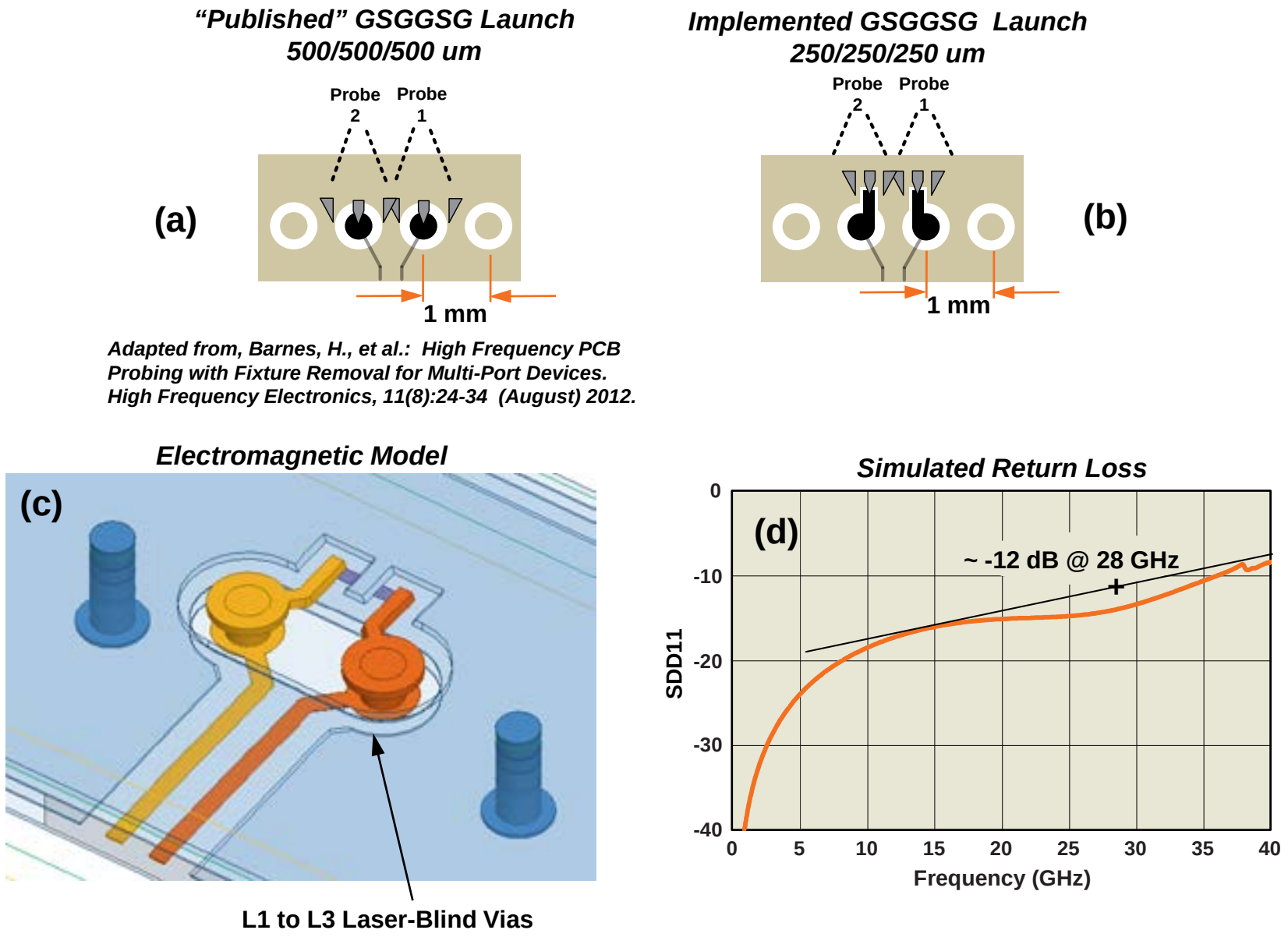

**Figure 3 – Probe launch design and simulated performance [45071]**

# General Channel Performance

The board design is intended to support short intra-board channels with stripline channel lengths of less than 10 inches. For that reason, the insertion loss is generally not of concern assuming that high performance low-loss laminates are used. Therefore we will primarily describe the return loss with respect to the electrical performance figure of merit. For $56G_{NRZ}$ the performance at or below 28 GHz will be discussed, which is often referred to the "Nyquist" frequency.

Next we will discuss the four stripline layers offering the best $56G_{NRZ}$ performance, to be referred to as the "control" cases to which other structures are compared. The via designs in these control cases are generally considered to be standard high data-rate designs, with the exception that the bottom of the vias terminate with a "padless" design. The padless approach is described in more detail in the next section.

We anticipated that the via designs with the shortest stub lengths and shortest via depths would have the highest performance. The L3 laser blind via was expected to have the best performance. The L12 blind via was not originally considered to be a top performing via since it is manufactured much like a microstrip and converted to a stripline during final lamination. L12 begins as a foil on an outer pre-preg layer (assuming foil-lamination construction commonly used for PCBs). It is then plated to fill the L1-12 via drill hole, and then L1 and L12 are patterned and etched to complete the top sub-laminate. The top sub-laminate is joined to the bottom sub-laminate with another sheet of pre-preg. Therefore L12 uses pre-pregs above and below to set the distance to the stripline reference planes. Pre-pregs are designed to conform to the adjacent layers during lamination whereby resin flows into regions where the metal has been etched away. In turn, we expected distances to reference planes to be much more variable in L12 than in a standard stripline layer that is adhered to a core laminate, which we assumed to be stable during a reflow process (although the standard stripline has a pre-preg layer on its other adjacent dielectric layer).

For various reasons, we specifically disallowed routing layers on the top and bottom microstrip layers. For example, board manufacturers often discourage the use of microstrip layers for high-speed routing.

In addition, for dense boards there are many component pads that would block the routing. The measured top and bottom microstrip performance was very satisfactory, to the extent that in the future we may consider using outer layers to overcome localized PCB routing congestion in a dense board. In a less-dense board, we have not ruled out outer layers for $56G_{NRZ}$ and similar applications.

Excluding the outer PCB layers from consideration, the next highest performance routing layer was expected to be on L10, which uses a blind via from L1 to L12, leaving approximately an 10 mil long stub. Similarly, L24 had the same stub length so we expected good performance on L24 but not as good as that from L10 since L24 used a larger 10 mil diameter via. Deeper vias are likely to be more affected by layer shifting.

Measured return losses, as depicted in Figure 4-b, are generally quite good, for all but the L24 routes. We are not aware of any existing $56G_{NRZ}$ channel specifications. Alternatively, we borrow from the OIF 28 Gb/s NRZ channel specification [6] and simply scale up the frequency for this specification from 28 to 56 GHz. This hypothetical specification for the return loss is superimposed on all of our return loss measurements. The L24 return loss does violate this specification beginning at approximately 25 GHz, and is quite likely the reason for the insertion loss dropout beginning at 25 GHz.

The measured insertion losses are shown in Figure 4-a. Note that some insertion loss anomalies exist. We can expect and confirm with simulation that very little insertion loss differences exist due to added loss along the via barrels. The via barrels for L24 are only about 6% of the total channel length. However, we did observe that L3 demonstrated noticeably lower insertion loss compared to the remaining channels, ignoring results above 20 GHz where via performance does begin to adversely affect insertion losses. We have been unable to identify a reason for this discrepancy. L12 and L24 do not use the exact same core and/or pre-preg; however, L10 and L3 were specified to have the exact same core and pre-preg materials and geometries. One might suspect that manufacturing width variations between L3 and L24 can cause insertion loss variability as a function of metal width change. However, of note is that all measurements showed return losses of approximately -30 dB at low frequencies, where the return losses at these frequencies are generally simply a function of the characteristic impedance match to the 50 ohm port impedances. This result suggests that all boards were fabricated across all layers to have an impedance tolerance of +/- 6% which translates to the -30 dB return losses.
.

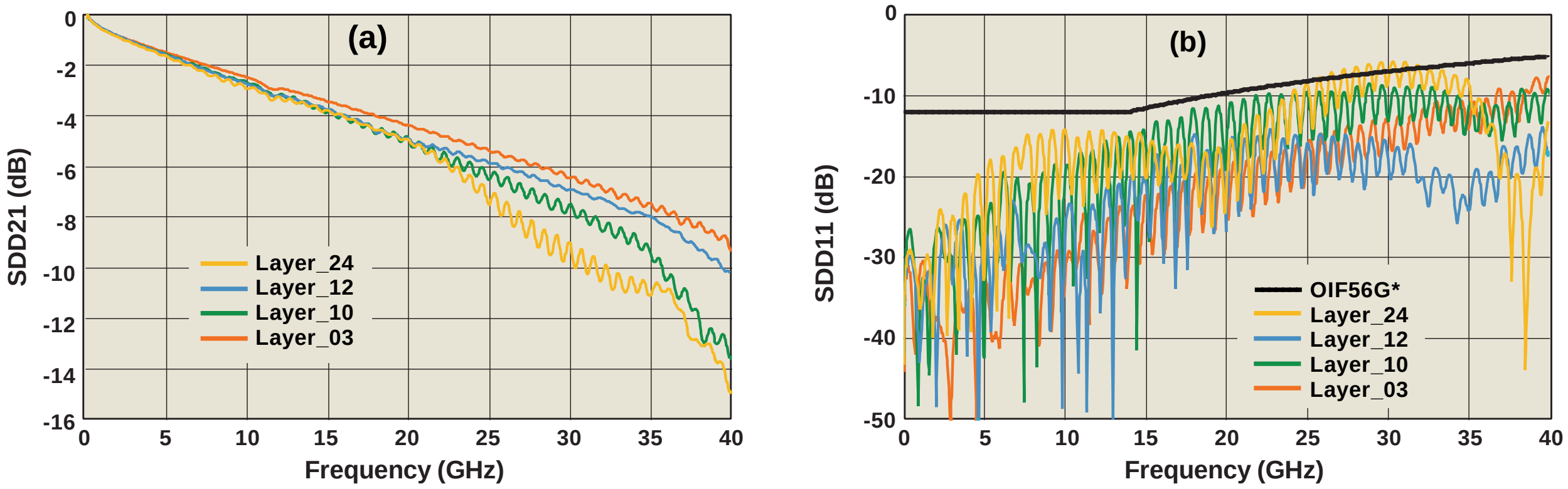

*Extrapolating OIFCEI-28G-MR normative channel return loss at 28.1 Gsym/s to 56 Gb/s NRZ encoding

**Figure 4 - Top four performing stripline layers – [45075]**

To better understand the performance of the vias, we also explore time domain reflectometry (TDR) responses from transient circuit simulation using the measured Touchstone files to represent the channel performance.

Figure 5 illustrates these TDR results for the L3, L10, L12, and L24 structures. The result for the bottom microstrip, L26, is also presented to help the reader understand the L24 response. The most basic observation is that the negative reflections for all of these vias demonstrate that they are overly-capacitive compared to a 50 ohm reference. Such a result was expected based on modeling. (We do not present or discuss modeled performances in this paper due to space constraints.) It is important to comment that the anisotropic behavior of dielectrics must be included in 3-D full-wave models of vias and related structures for the accuracy required in $56G_{NRZ}$ applications. Anisotropic behavior for PCB materials has been discussed in the past [5,7].

A number of methods exist to reduce the excess capacitance of a via, some of which are practical and others of which are not. Traditionally, the best way to improve the via performance is to reduce the via stub, by way of back-drilling. However, the worst-case via stub for the top four performing layers spans only about 10 mils (through two dielectric layers) which generally cannot be back-drilled given today's typical back-drill tolerances. The next most effective way to reduce via capacitance is to increase the antipad sizes. We previously noted the use of a larger antipad on the metal-flooded top layer. The other top-laminate layers must also support significant power delivery current, so removing additional metal by increasing antipad size is often not permitted. However, the bottom-laminate metal layers typically offer little power delivery resistance reduction and therefore larger antipads can be considered for these lower layers. The L24 via will require the most performance improvement and the electrical length for that via barrel is very close to the bit period for $56G_{NRZ}$. Therefore, it is likely that much of the via barrel must use the larger antipads to realize a performance improvement. Stated alternately, a high capacitance region along a via barrel should be relegated to a small fraction of the via length. However, it is prudent to use simulation to gain improved insight into these variations since the via is a complex 3-D structure.

Other design variations should also be considered to improve via performance that we have not had a chance to explore. Vias for the positive and negative (P/N) polarity signals comprising a differential signal pair can be spread apart slightly to reduce their differential coupling. We have been reluctant to use wider antipads, which constrict current delivery, but lengthening antipads can reduce via capacitance, which should create minimal plane resistance increase. Via stub capacitance can be reduced by greatly increasing the antipad sizes for those layers surrounding the via stub.

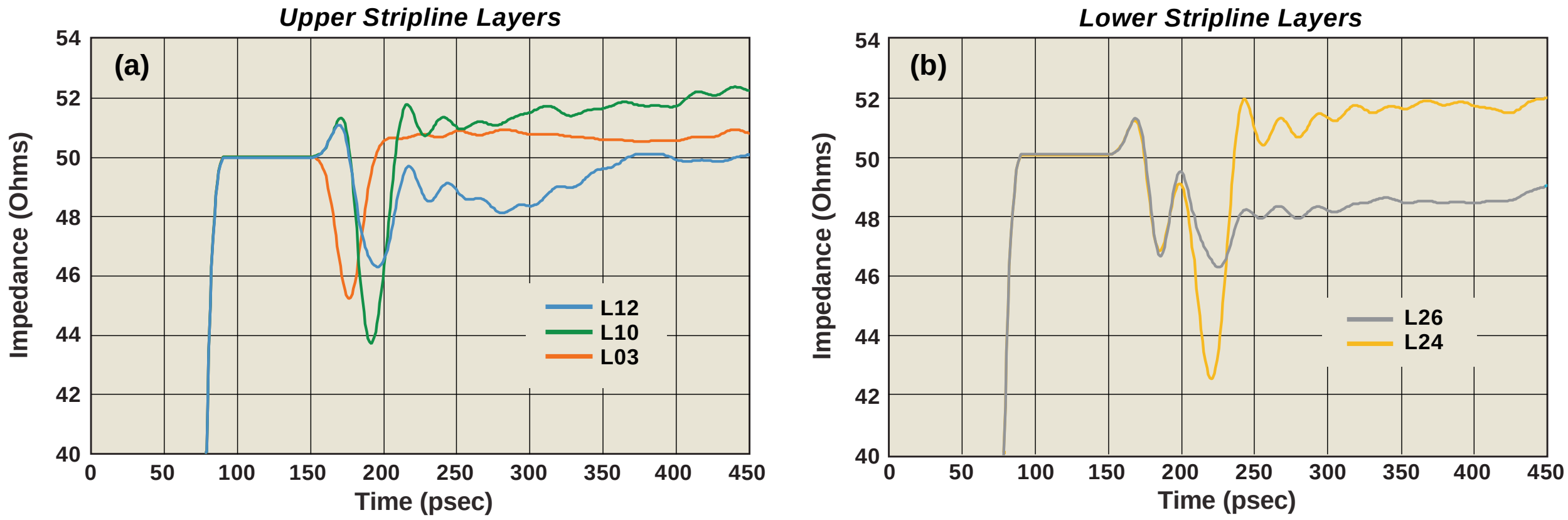

**Figure 5 - Typical TDR response of L3, L10, L12, L24, and L26 – [45076]**

## Alternative Via Designs

For very high SerDes data rates, pin-field/via design strategies of stub-length control and antipad sizing are not always sufficient. Therefore, in this section, we continue to discuss a number of other design strategies to further improve electrical performance.

It can often be advantageous to work closely with the PCB manufacturers to understand their fabrication processes. We had some prior experience with “padless” vias, and requested that same capability from our manufacturer. The manufacturer generally requires larger pads on internal layers so that they can drill through these pads to form the mechanically-drilled via. However, the outer layers during the drilling process are non-patterned copper foil. After drill and via barrel plating the outer layers are then patterned and etched to form the via pads, routes, and other features. Therefore, outer layer pads can be more precisely defined and need to be oversized from the drill diameter by the amount of drill position accuracy. The board manufacturer allowed the use of a 1 mil annular ring on the outer layer pads which is considerably smaller than the 5 mil annular ring required for inner layer pads. For simplicity, we refer to this via with a much smaller outer layer pad as a “padless” via even though these pads still have a small annular ring.

In a typical design, the padless via cannot be invoked at the BGA launch since a prescribed pad size is required for robust solder assembly. For our test boards, the option was available to use the padless vias not only on the bottom layer but also on the “intermediate” outer layers on L12 and L13, since these layers are top and bottom layers during individual sub-laminate processing.

As noted previously, padless via design was selected as the standard design process in these test boards. Therefore all of the via designs used padless vias except for special cases which will now be described. Since it was desired to emulate PCB channels driven with top mounted BGA devices we only applied the padless via on layers where the mechanical drill terminates, i.e., the “terminating” pad. Measurements presented in Figure 6 compare the performance of the standard padless via to the use of a terminating pad that is sized to the diameter used for the inner pads. These results demonstrate that the padless via is essential for enabling the L10 via to meet $56G_{NRZ}$ performance requirements as the return loss changes from -9 to -5 dB at 28 GHz. The L24 via with the full-sized bottom pad exceeds the specification at approximately 11 GHz, but both vias appear to have roughly the same performance at 28 GHz. We believe that this result is an artifact of the “dB saturation” effect, by which is meant that the performance is already poor, roughly -5 dB at 28 GHz, so it is difficult for a full-sized bottom pad to make the performance noticeably worse. With some of the design modifications discussed above, it is possible that the L24 via can be designed to have more robust performance for $56G_{NRZ}$ applications and would therefore require or benefit substantially from the padless via.

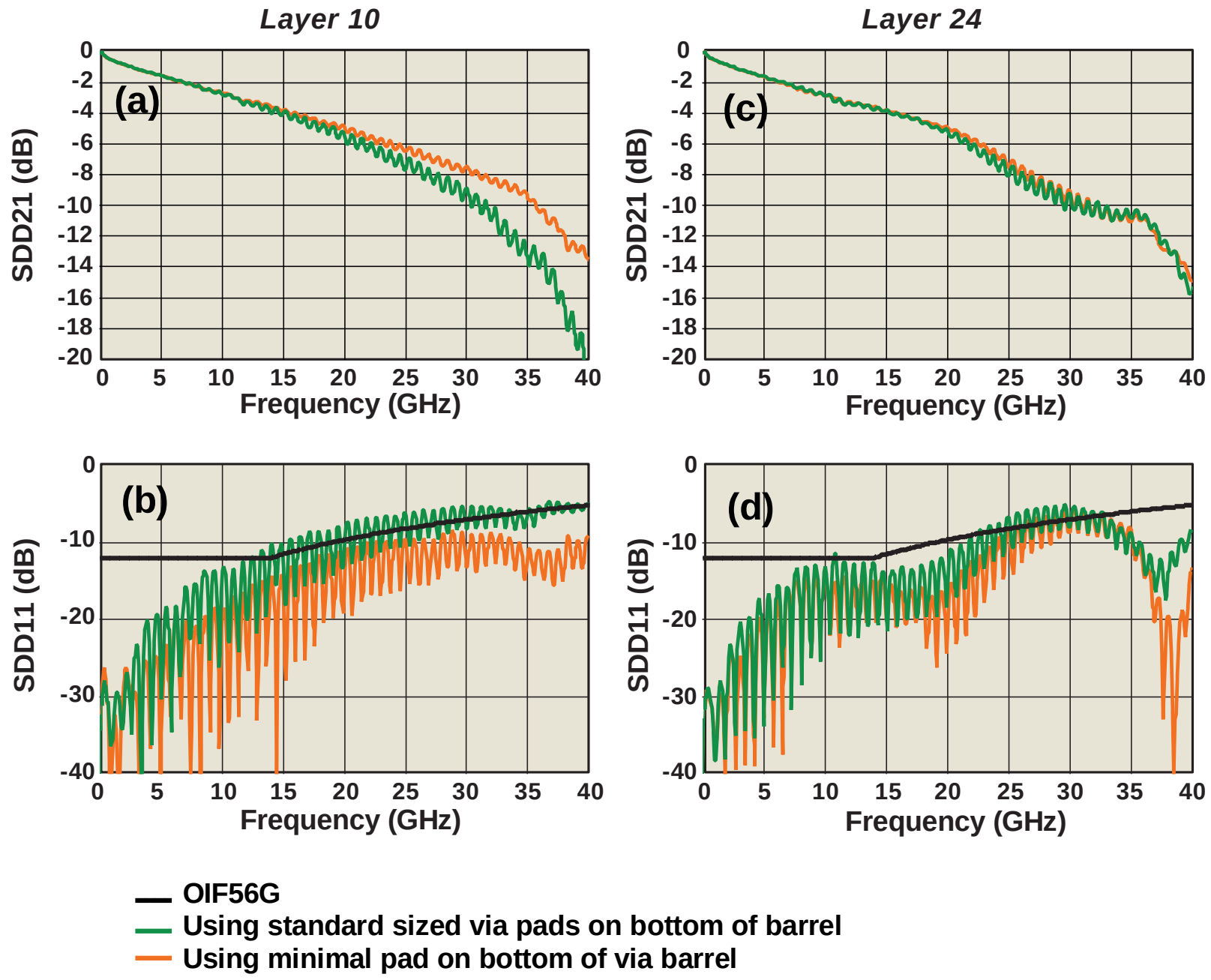

**Figure 6 - Performance lost when using standard sized bottom pad. [45077]**

Next we consider the effect of different break-out styles (as examples, see [8], page 13). To best isolate the effects of the break-out styles, the L3 signal layer was used to minimize any other perturbations that might arise from vias to deeper signal layers. The right side of these structures is shown in Figure 8-c,d,e. The control structures use broadside breakout. The left side of these structures (not shown) also uses the broadside (control) breakout. Dense BGA devices typically use in-line differential signal pair assignments. In-line signal pairs typically use standard in-line break-out; alternately, back-jogs may also be considered. Break-outs that have in-line pin assignments inherently have 1 mm of skew (for devices with 1 mm pitch pins). The 1 mm of length skew results in 5.9 psec of timing skew, which is typically corrected by jog-out length tuning. For more direct comparisons the jog structures are considered part of the pin-field structures. Typically one or more jog-outs are used for length tuning.

In Figure 7-a, we present the measured performances of the control channel versus channels employing back-jog and traditional in-line break-outs. The in-line approach utilizes jog-outs to length-match the differential pair, whereas the back-jog design incorporates length tuning using a predetermined routing pattern within the pin-field. The in-line design shows little performance degradation whereas the back-jog has small but noticeable additional insertion loss at 28 GHz. Oddly, the back-jog does not have significantly higher return loss than that of the in-line structure.

Next we compare the performance of the control structure to that of the “neck-down” structures, as shown in Figure 8-e. The primary benefit for these neck-down structures is to achieve two routes per routing channel by reducing the line width and decreasing the line spacing to achieve a 100 ohm-differential impedance. It was very difficult to discern performance differences of either neck-down design from that of the control channel. This result may prove significant, since we had previously believed that neck-down structures would have worse performance than structures that did not employ neck-downs.

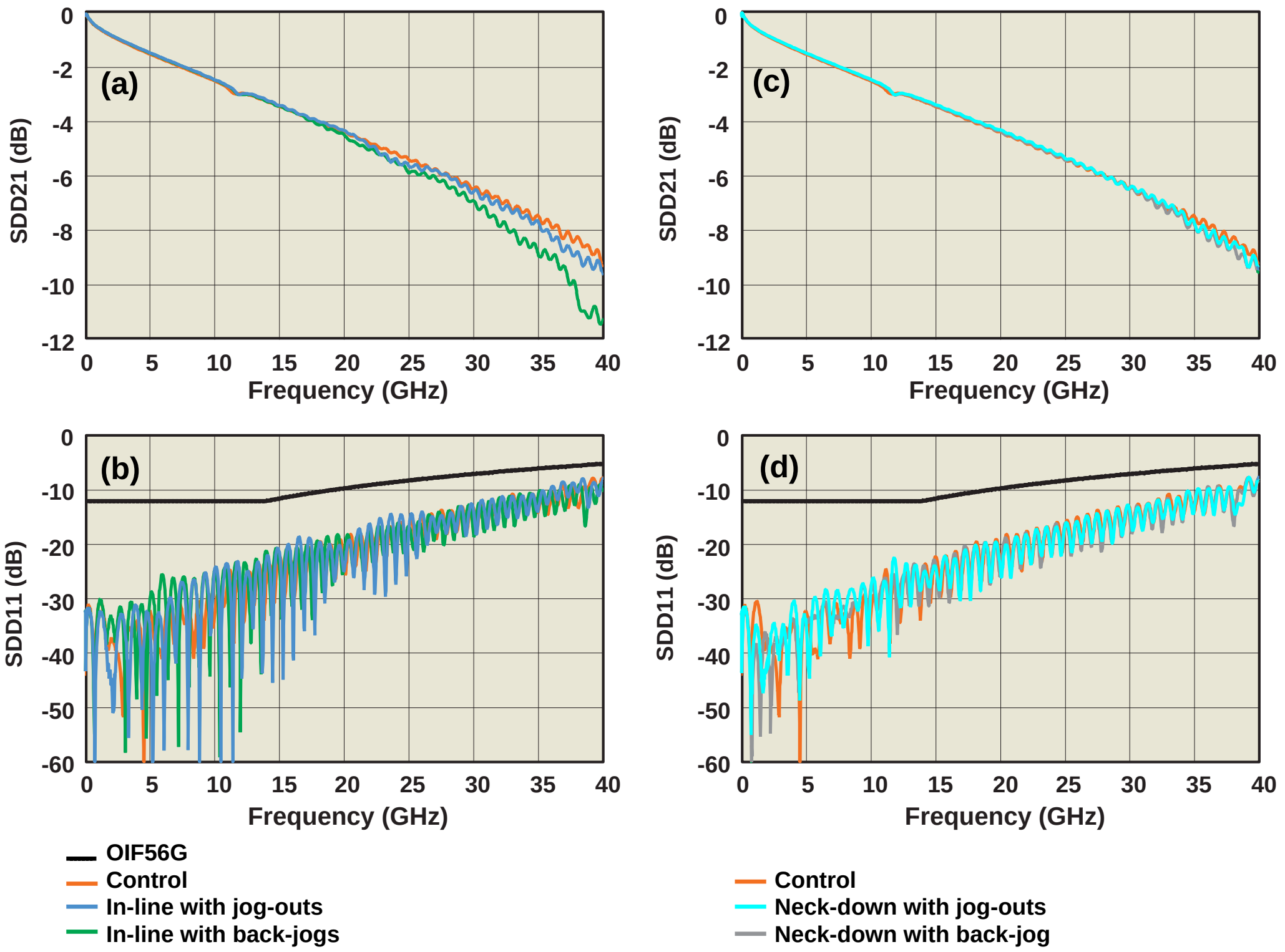

**Figure 7 - IL and RL using differing pin-field escapes and neck-down routing [left] a) control, b) back-jog, c) jog-out, [right] a) control, b) back-jog with neck-down, c) neck-down jog-out [45078]**

In Figure 8-a,b we present residual channel skew of the structures employing the differing break-out designs. We found that the balanced broad-side break-out has essentially no skew, whereas all other break-outs have approximately 1 psec of skew for $56G_{NRZ}$ applications. Later we will make the case that the channel skew budget is only 3.6 psec and therefore adding 1 psec per break-out, with two break-outs per channel, consumes over one-half of the channel skew budget, assuming that the right side of the channel uses the same break-out style as that used on the left side.

In order to de-skew residual break-out/jog-out skew, one might be tempted to add about 0.17 mm of stripline, i.e., the electrical equivalent of 1 psec, to the stripline with the jog-out; however, this methodology becomes error-prone and logistically difficult with adjusting PCB design rules to recognize pre-skewed lines.

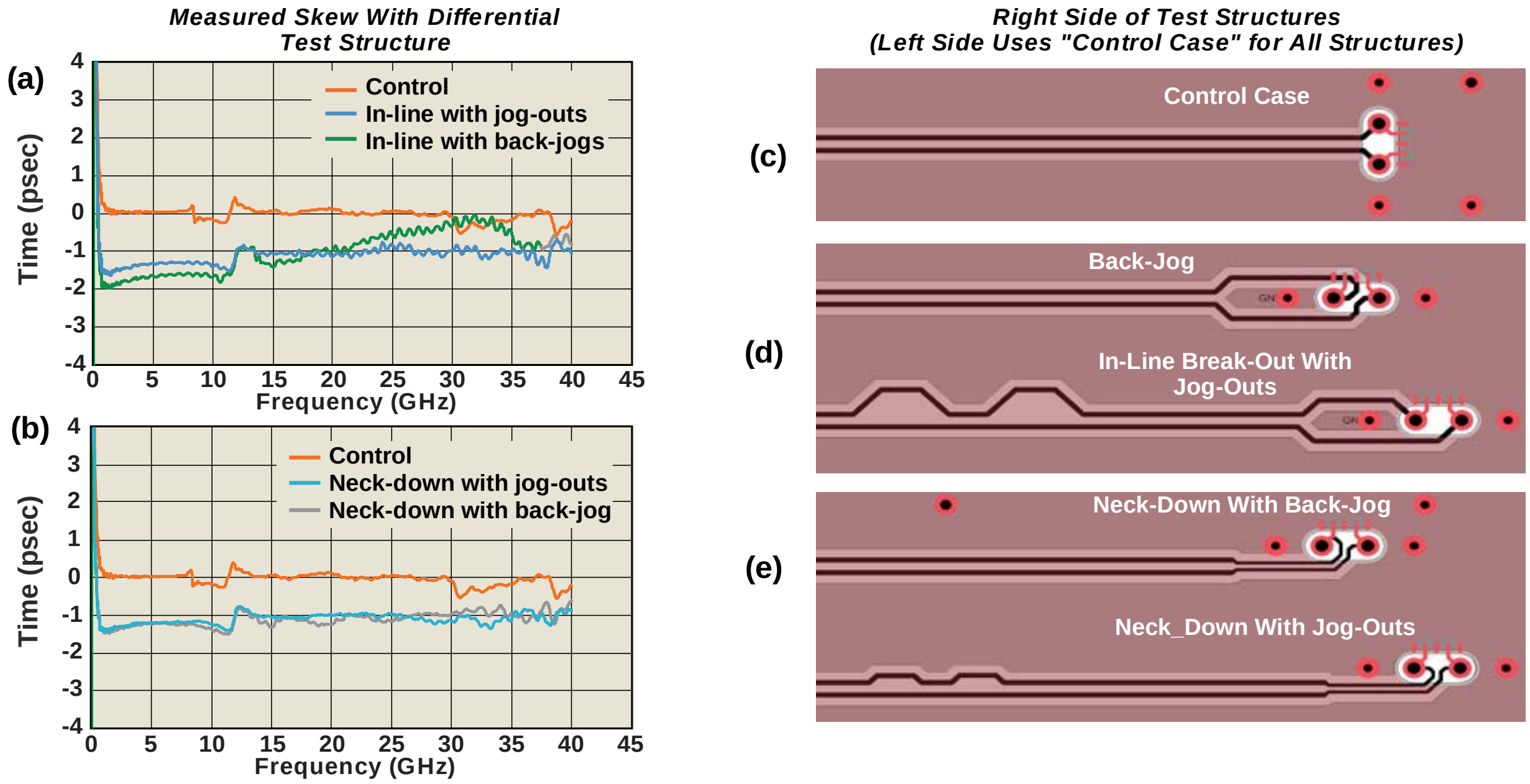

**Figure 8 - Channel skew for (a) control line, (b) back-jog, (c) in-line breakout with jog-outs, (e) neck-down and unbalanced back-jog, (d) neck-down and jog-outs [45085]**

Next we study the effect of channel performance when using jog-outs for tightly- versus loosely-coupled striplines. Starting with the L3 control structure having 5 mil width and 12 mil pitch striplines, two small jog-outs were incorporated, to add 1 mm of length skew to the P-stripline near the left probe site. Then the same jog-out patterns were added to the N-stripline to balance the channel length skew at the right probe site. The probe patterns are moved closer by 1 mm so that these test structures match the total physical 4.2" length of the control structure. The measured performance is compared to the control structure in Figure 9-a,b. Note that the jog-outs did cause a small insertion loss ripple as well as some noticeable return loss increase when compared to that of the control structure.

We selected a 4.1 mil wide line and an 8 mil line pitch for the tightly-coupled stripline case. The jog-outs were implemented exactly the same as in the loosely-coupled case. In Figure 9-c,d measurements show that the higher single-ended impedance of the tightly-coupled jog-outs caused excessive insertion loss drop-out, and caused the return loss to nearly violate the 56G specification. In this case, the control structure used the tightly coupled line width/spacing definition. With these results it seems apparent that tightly-coupled striplines for $56G_{NRZ}$ should only be considered if no jog-outs would be required. Alternatively, if tightly coupled lines are used to implement neck-downs, then jog-outs could simply be added after the neck-downs transition to the standard loosely coupled striplines.

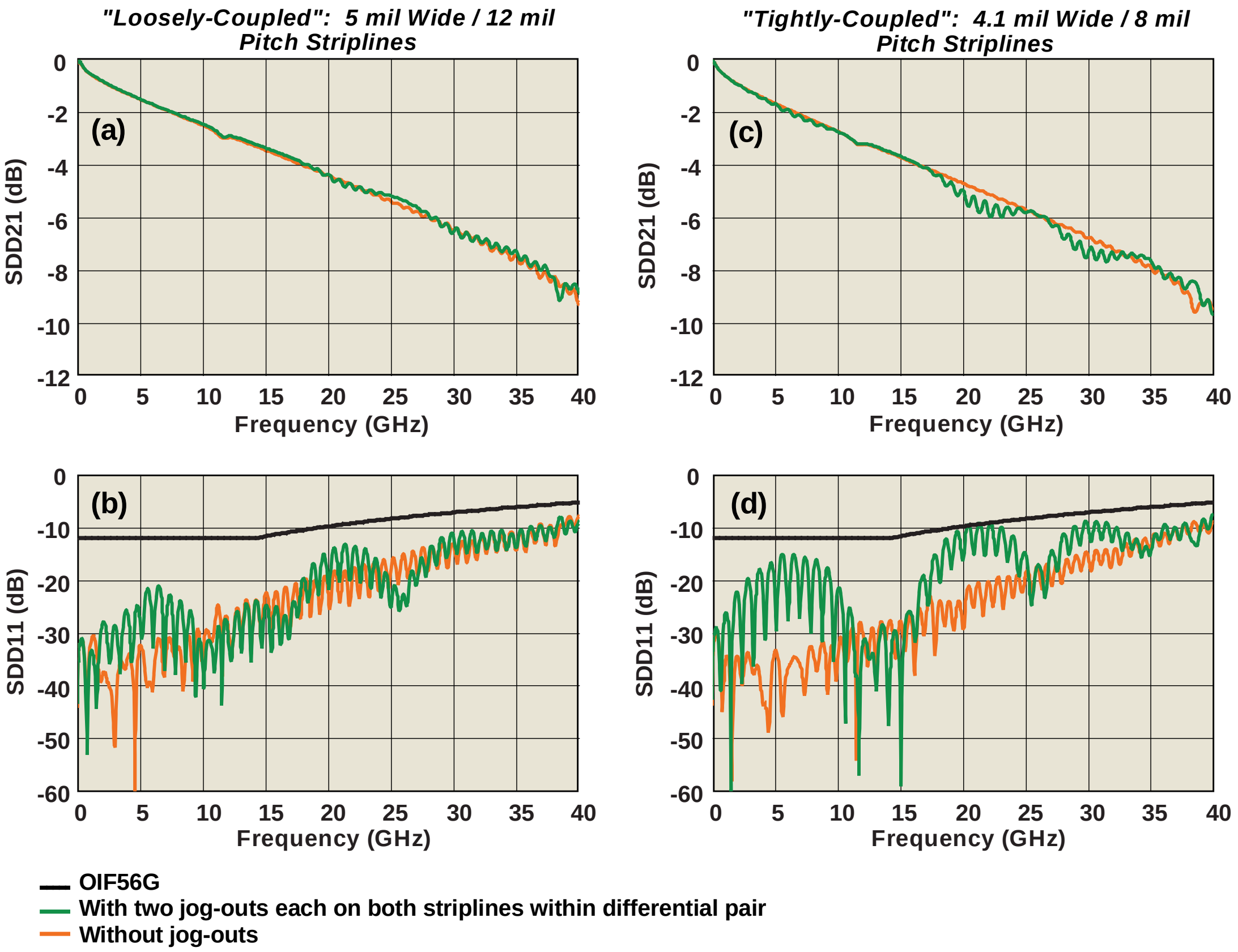

**Figure 9 - Effects of jog-outs for both loosely and tightly coupled striplines. [45080]**

The obvious significant benefit of blind vias is that there are no via stubs which significantly limit electrical performance. However, the connect pad, i.e., the pad from which the stripline exits, can often be of substantial size. Even some laser blind vias might be large enough to affect $56G_{NRZ}$ signal performance depending on the implemented technology and via depth into the board. We investigated how to improve these blind vias by voiding the plane(s) above or below the connect pad. For example, for the L3 laser blind via, the via extends from the top layer to L3, so we voided the L4 and L5 layers beneath the L3 pad. Voiding multiple planes usually has limited performance benefit compared to voiding only the nearest plane – except when the planes are very close to the connect pad. In this case the L4/L5 spacing was only approximately 1 mil, since a BC core was implemented beneath L3 to assure reasonable power integrity.

Figure 10-a,b illustrates some measured performance improvement for the L3 blind via in adding the plane voids, compared to the control case where these voids were not used. The return loss performance improvement up to 35 GHz was substantial. However, typically significant link performance improvement is not observed by enhancing a channel that already easily meets specifications. In another case, with results shown in Figure 10-c,d, vias extend from top to L12, then extend 2.54 mm on layer 12; and then blind vias extend back up to L10. This use of two vias avoids the via stub and nearly meets the 56G specification without further modification. We can additionally void the L9 and L8 planes above the L10 blind vias to achieve even better performance. This resulting channel performance matches quite closely to the control channel using the L10 PTH via (which has the 2-layer via stub to L12). Therefore, there is no need to use the extra vias since it is more complicated to do so unless special routing considerations such as bypassing blocked paths arise.

Note also that the plane voids that extend beyond the via drill layers had to be added manually due to a limitation of our PCB layout software. From a pragmatic perspective, it is suboptimal to implement any design changes manually since all subsequent verification must be performed manually. Further, any later design changes would require moving the location of the manually placed voids. Such a process is fraught with error - if the via designs are not implemented correctly then disastrous results will occur for $56G_{NRZ}$.

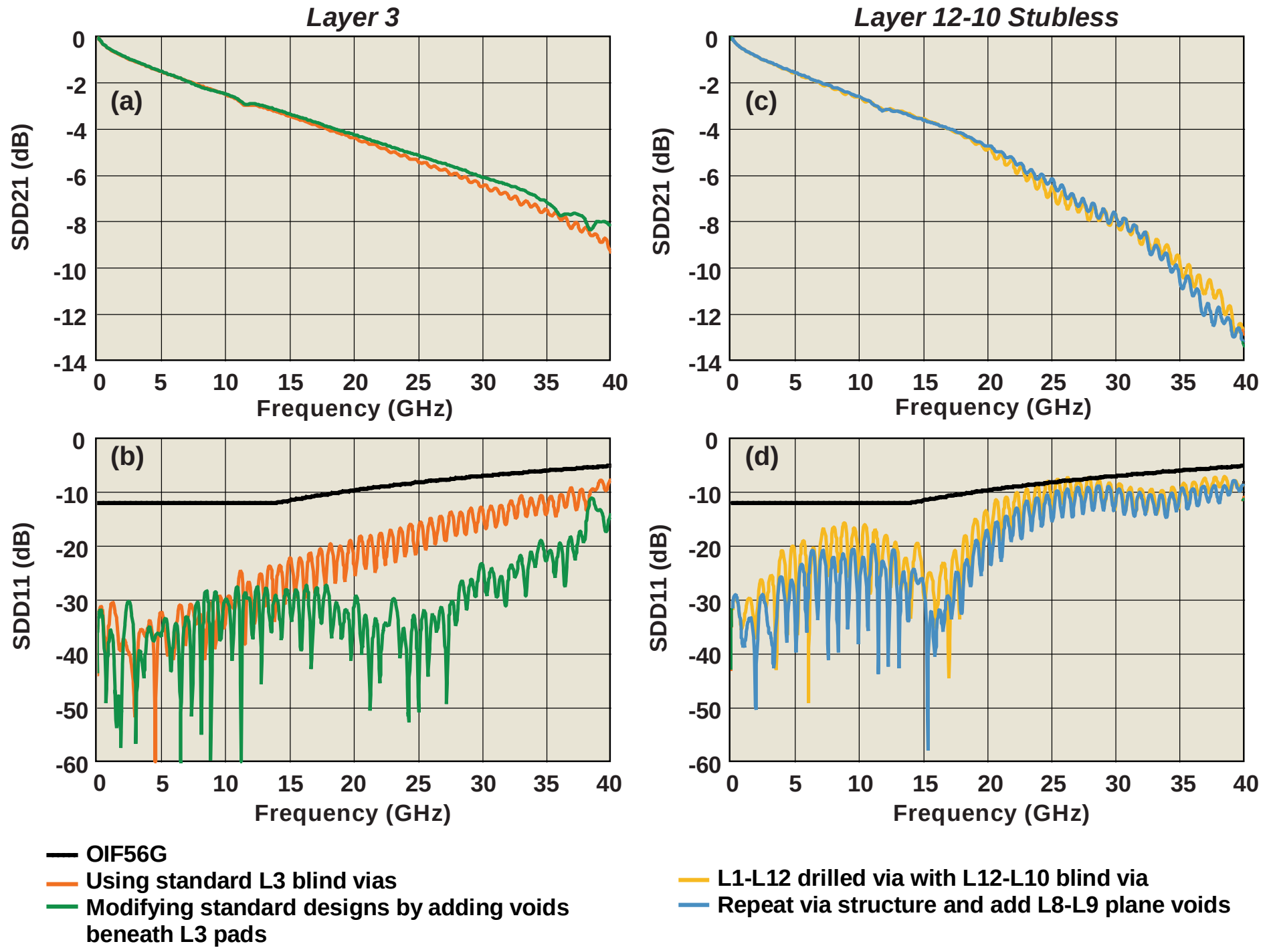

**Figure 10 - Improving performance by voiding reference plane closest to blind via [45081]**

When presented with a range of via structure types it is of interest to consider whether it is possible to use multiple vias to break-out on a targeted stripline layer, whereby the use of the multiple vias would result in better electrical performance than that of the single via. We constructed a number of test cases, the measured results of which appear in Figure 11. In Figure 11-a,b, for completeness, we repeat data previously discussed above for the L10 case. We found that using the L12 PTH followed by the L12-L10 blind via actually performed slightly worse, just pressing up to the 56G specification, than using only the L10 via with a short stub to L12.

Using multiple vias to eliminate the stub for routes on L15 can improve performance substantially, as is illustrated in Figure 11-c,d. There are two approaches to using multiple vias to eliminate the L15 via stub. First, one can via to the bottom, use a short microstrip length to a bottom sub-laminate PTH via, and exit onto L15 from that second via. This approach leaves a short residual via stub from L15 to L13. A second approach is to via to the bottom, use a short microstrip length to a bottom sub-laminate PTH via, transition back to L13, then a short L13 route to a L13-L15 blind via. This 3-via approach appears to be novel, by eliminating all via stubs, but actually provides slightly worse performance than that measured with the 2-via approach.

In Figure 11-e,f the measured performance for the L24 standard via versus that of a “stubless” version is shown. For the latter case, vias pass to the bottom layer, followed by a short-distance routing on the bottom microstrip to use a L26-L24 blind via to L24. The stubless via approach does not appear to offer much performance advantage.

The measurement results on these stubless vias do not seem particularly surprising. The approach works very well if a very long via stub is bypassed. However, often the use of multiple vias to reduce or eliminate short via stubs can actually decrease via performance slightly. This result underscores some of the interesting cost benefits of having a multitude of via technologies available within the same board. For example, we used 2.54 mm of microstrip etch stitching between multiple vias. Perhaps better performing vias could have been achieved by co-designing these multiple vias that are placed adjacent to one another. It is very difficult to design and fabricate the many variations of vias used, their void sizes, and via spacings to attempt to cover the breadth of design choices.

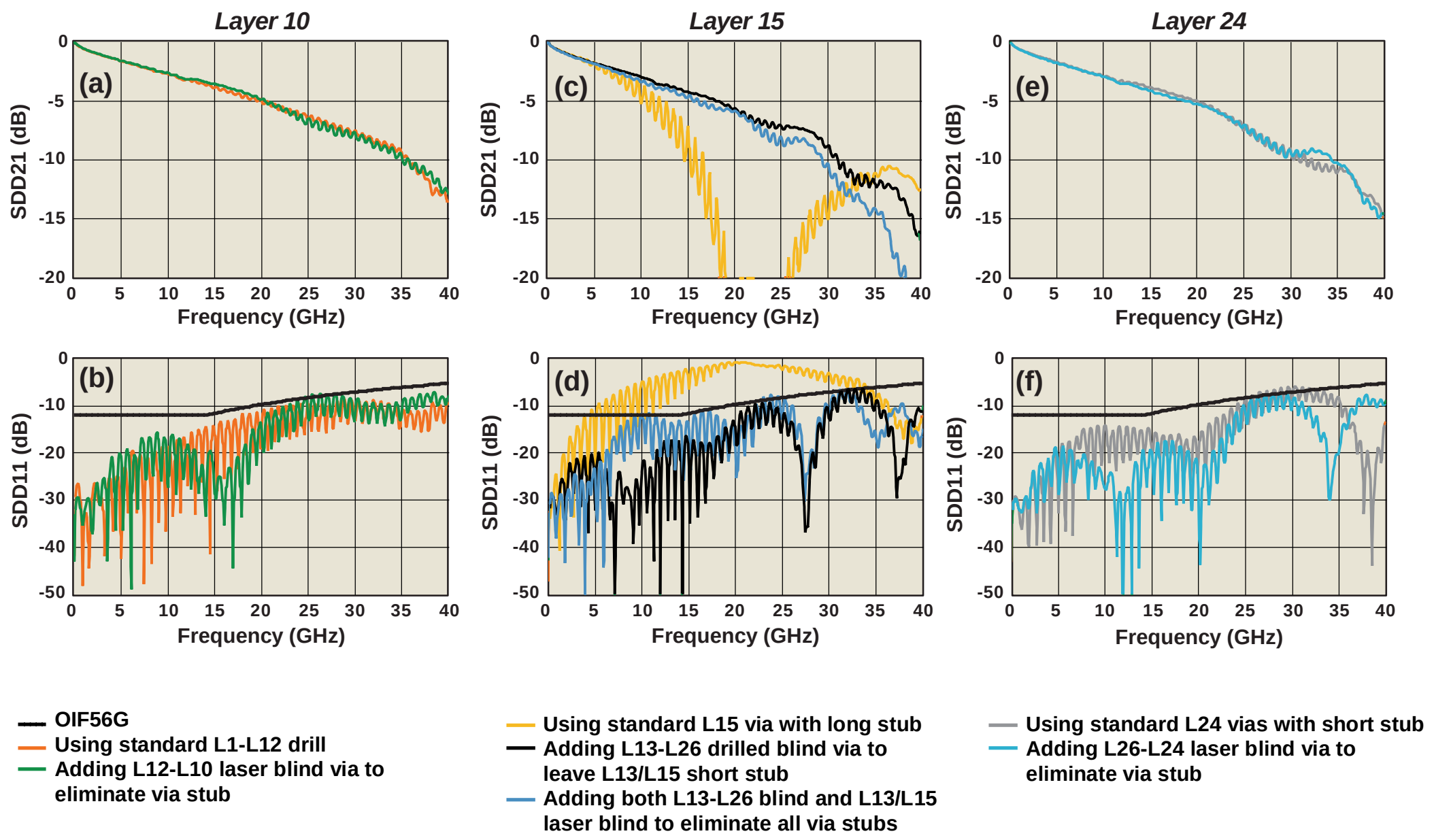

**Figure 11 - Using multiple vias to remove via stubs [45082]**

# Impedance Control Study

A concern that many engineers may have is that at higher data rates stripline impedance control may have to improve beyond the typical industry standard of +/- 10%. We do not have significant experience in tracking impedance variation across a board. We did closely monitor the impedance of microstrip launches on a different test board, and the results were alarming, with insufficient line width control and high concomitant impedance variations [5]. In addition, we have been informed by various PCB manufacturers that the added top-layer manufacturing steps and the fact that the microstrips typically reside on top of a pre-preg layer can both cause microstrip impedance variability along its length. Based on measured results, our test boards did not appear to suffer from significant impedance variation on traditional microstrip layers (L1 and L26) or on L12 with microstrip-like features.

In Figure 12-a and b are shown the TDR response and return losses, respectively, across five of the control L3 striplines, each identical structures, but from different boards. The TDR results show only small deviations along the length of these striplines, perhaps up to +/- 4% (or +/- 2 ohms), but many striplines have even better control. Such uniform impedance control was a surprise, and thus we attempted to determine the performance impact if greater impedance deviation had been encountered.

We employed simulation to investigate the effect of +/- 10 % impedance variation. It is our understanding that PCB manufacturers typically average the impedance over the length of a TDR response and if so, some localized variations would exceed the typical specifications. As illustrated in Figure 13-a, we examined the electrical performance, running 31 Monte-Carlo cases for a 10” stripline in which the impedance was allowed to vary uniformly from 45 to 55 ohms. This result did not provide much information, other than to show expected worst-case results to compare with other simulations. In Figure 13-b, the 10” stripline was modeled as 10 segments, each having independently random length and impedance from a uniform distribution ranging from 45 to 55 ohms. Therefore, it is possible that adjacent stripline segments have the worst impedance discontinuity, e.g., a 45 ohm segment adjacent to a 55 ohm segment, and it is also possible that adjacent segments have good impedance matching. This model is likely to be quite pessimistic as it allows for impedance ranges from 45 to 55 ohms. In practice, it is likely that the average impedance could be high or low, and that the variation from the average impedance might be much smaller than +/- 10 %. The Monte-Carlo simulations demonstrated that even with this hypothetical pathologic worst-case, the random variation actually resulted in low overall performance spread, especially with the small variation of the insertion loss.

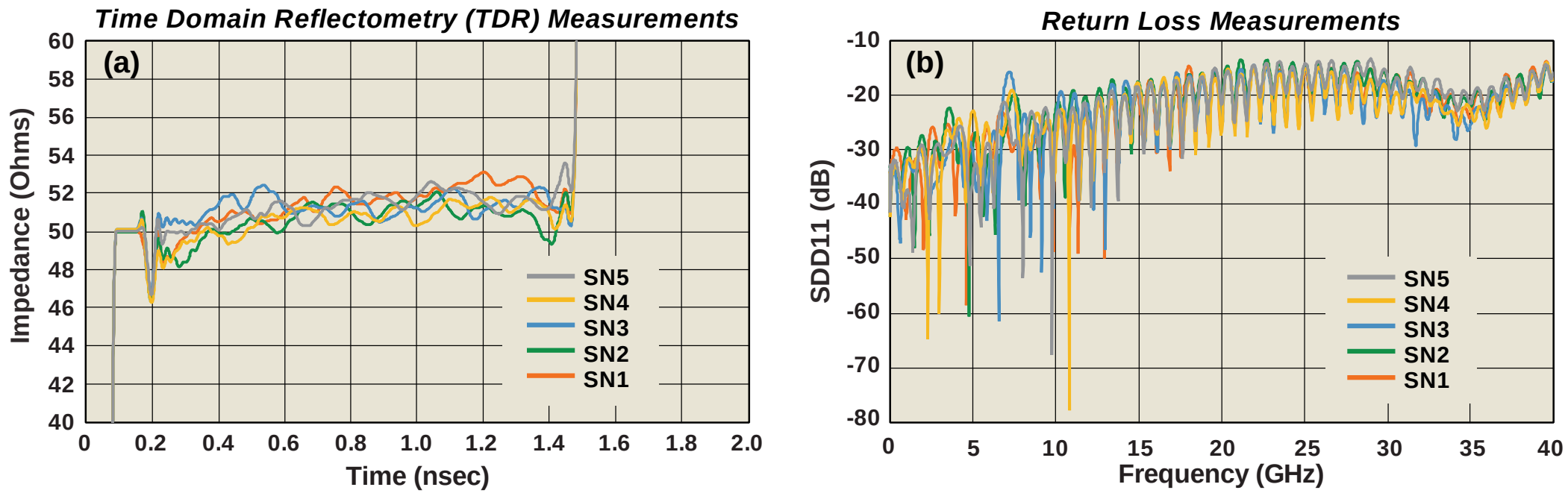

**Figure 12 – Impedance and return loss variation of Layer 12 [45083]**

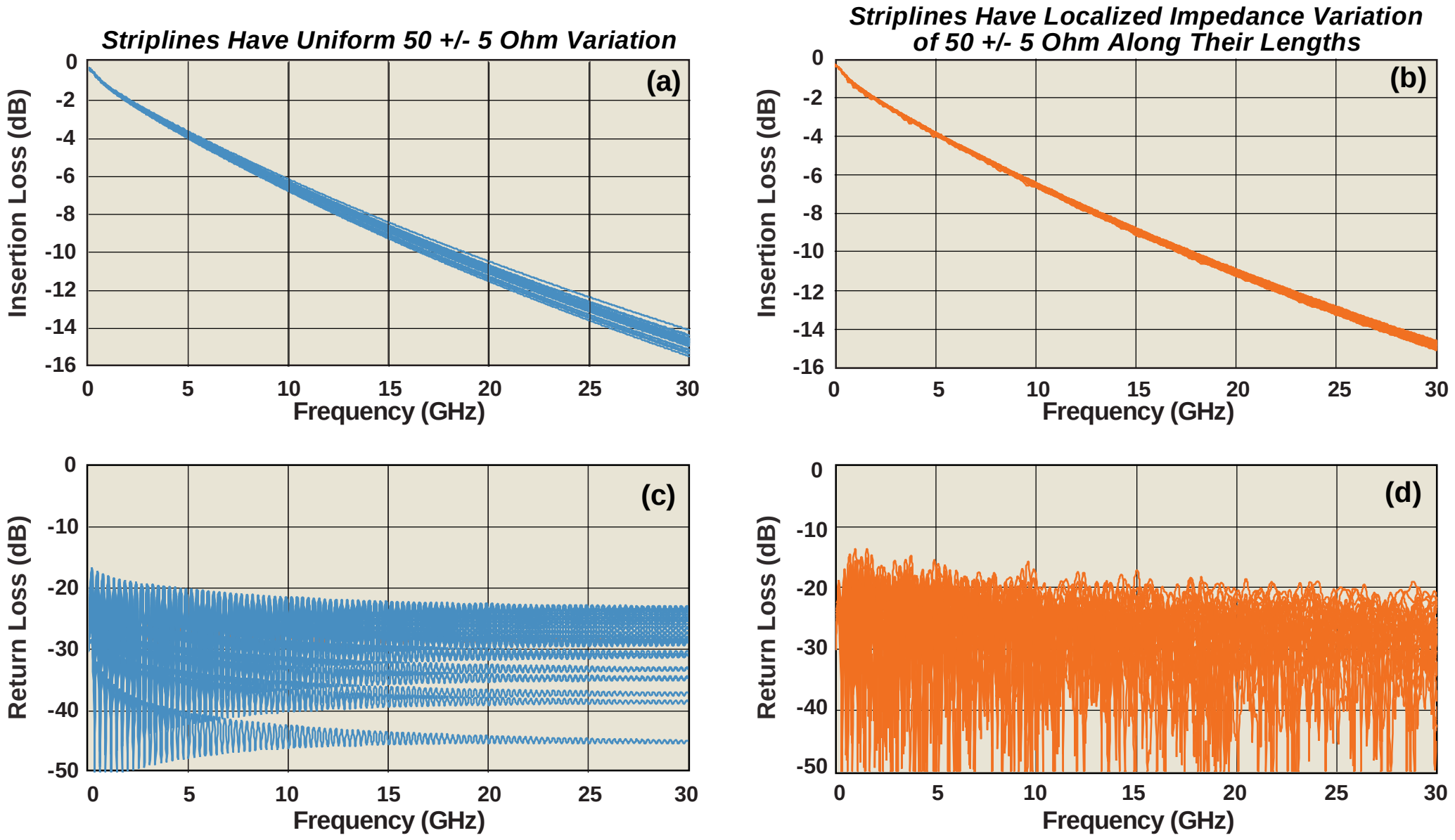

**Figure 13 – Effect of impedance variation within +/- 10% (a,c) uniform impedance and (b,d) random impedance across stripline comprised of 10 segments having random impedance [45084]**

# PCB Layer Shifting Implications

One of the concerns with PCB technology employed at very high data rates is that the layers comprising the PCB are not perfectly aligned after processing the board. This so-called layer-to-layer alignment can be caused by several factors. Generally, a PCB is comprised of metal-clad cores separated by pre-preg. The cores use fully-cured reinforced resin, while the pre-preg utilizes partially-cured reinforced resin. During the laminate pressing, the pre-preg becomes viscous and flows into voids where the metal has been etched away on the cores, “glues” the cores together, and then becomes fully cured. During this process the metal layers can become misaligned.

It seems likely that the effect of layer shifting within a PCB is to increase capacitance in the vias, assuming that antipads are nominally centered around the via. As we noted above, vias within a 1 mm pitch pin-field are already nominally overly capacitive, even with the lower dielectrics offered by the latest high-performance material sets. However, it is very difficult to estimate the via performance degradation due to layer shifting. It is not practical to sample structures on many boards to acquire a measured statistical basis for the effects of layer shifting. We have attempted to measure the layer shifting on a few boards and then discuss the results with the board manufacturer, in the hope of achieving an understanding of the nature of the layer shifting and related tolerances.

Figure 14-a – presents the design of such a structure to measure the layer Y-displacement from L1 to another layer “L_N”. Using x-ray imaging we inspected which metal strips align. Since these strips have as-designed incremental misalignments between the top and test layers it was possible to determine quickly the offset between the top and L_N test layer. As shown in Figure 14-b, these structures (rotated 90 degrees) repeat across all layers, maintaining a reference on the remaining top layer of the 26 layer test board. A copy of this structure is rotated and placed on the board to capture both x and y layer shifts.

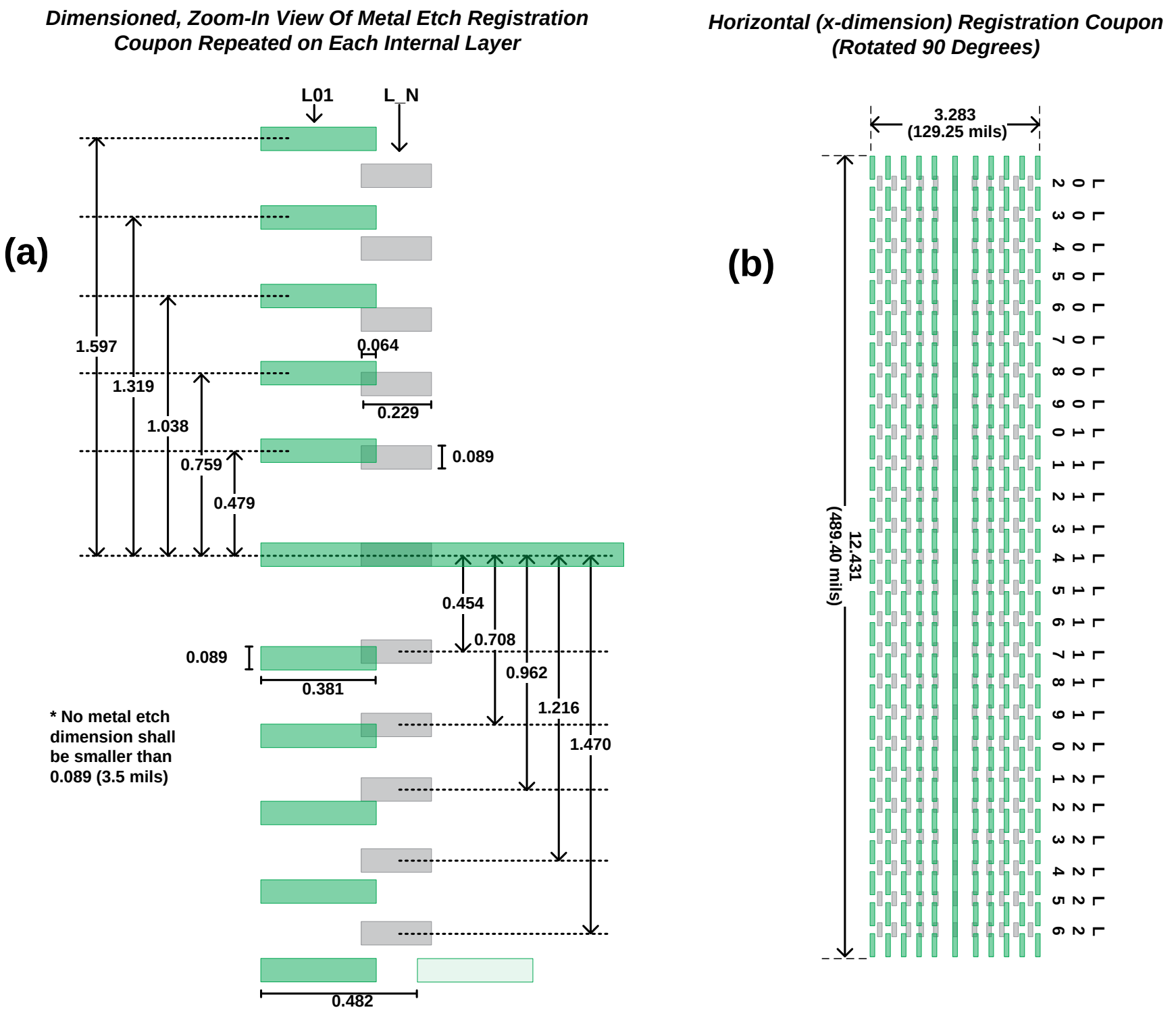

**Figure 14 - Layer registration test structures [44903]**

In 2007 we used these comb registration structures on a test board as shown in Figure 15-a. This board covered the entire useable area of an 18”x24” panel, and we used a material set having one of the lowest electrical losses among available materials at that time. This board consisted of eight layers, with thicker internal cores used to increase the board thickness to approximately 100 mils. Six registration sites (as circled) were distributed uniformly over the panel. Several of these test boards were manufactured; vector plots representing layer shifting are presented in Figure 15-b,c, and d. The shifts of each of the lower seven layers in relation to the top layer are represented both in magnitude and direction. For example, the shifting on Panel #4 was especially high, with Layer 5 shifting almost +/- 7 mils across the panel, a significant example illustrating that the term “shift” is somewhat misleading. In this case the Layer 5 exhibited a high degree of *stretching* – essentially stretching out fairly uniformly and elongating by 14 mils across the panel. The performance of the vias was very poor compared to that expected by simulation, and it was very likely that this observed layer shifting caused the electrical performance to deviate so far from the simulated performance of the vias.

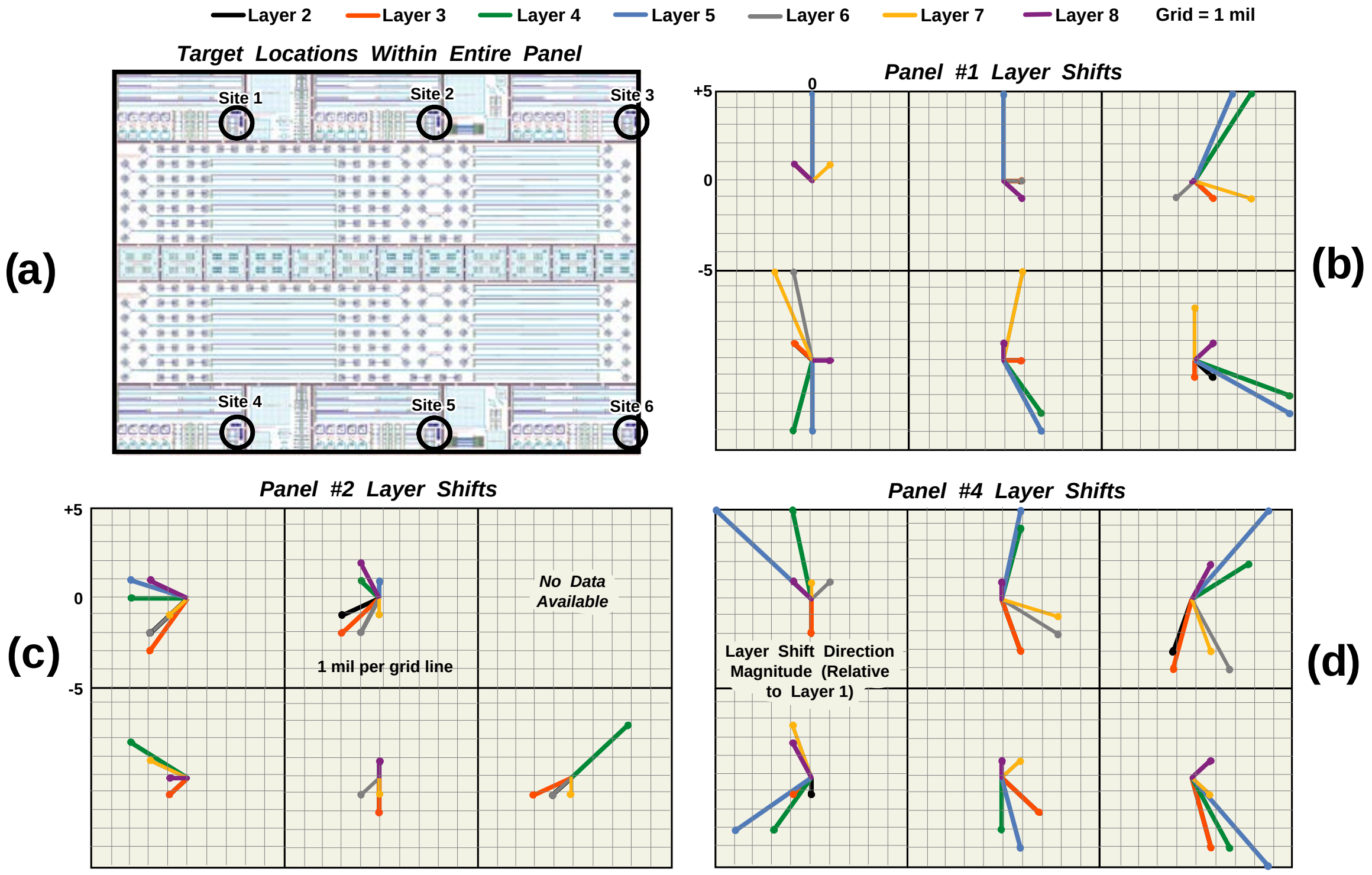

**Figure 15 - Layer to layer registration measured on 2007 test board at six sites across three panels [23141]**

Given this past experience with very high layer shifting when using materials for high-speed applications, we were surprised and pleased to observe very little layer shifting on the test boards described in this manuscript. These layer shifts are summarized in Table 1 for SN1. We spot-checked our additional boards and found the shifts to be similar to the material presented in this table. It should be noted that the layer shifts presented here are from only two test sites on the reference board, whereas for the board fabricated in 2007 and described above, layer shifts were observed essentially across an entire 18”x24” panel. For these newer boards, we detected a maximum shift of only 2.0 mils.

Measured layer shifts on these most recent test boards result in other interesting observations. Note also that we use layer shift tolerances for more than the design of the vias. For example, knowledge of the amounts of inter-layer shift provide PCB design guidance on how close a stripline can be from plane edges, such as between the rows of antipads in a BGA pin-field. In these cases, depending on stripline placement and shift tolerances, a reference plane could shift completely away from the stripline, yielding a buried microstrip structure. Even worse, it may be possible for both layers to shift away from the stripline, resulting in a significant impedance change and the potential for substantial crosstalk from similar unshielded striplines in the signal layers above and below the victim stripline. Formerly, we had assumed that layers on the same core do not shift much with respect to each other, since the core should be fully cured. Further, advanced PCB manufacturers use laser-direct-imaging to define the metal patterns, such that tolerances for position accuracy between top and bottom sides of the core *should be* 1 mil or less. A review of previous layer registration measurements has not yielded data that can express any certainty that intra-core metals should inherently have similarly low layer-to-layer shifting.

| | ***South West Test Structure*** | | | ***North East Test Structure*** | | |
|---|---|---|---|---|---|---|
| ***Layer*** | ***x-offset (mils)*** | ***y-offset (mils)*** | ***absolute-offset (mils)*** | ***x-offset (mils)*** | ***y-offset (mils)*** | ***absolute-offset (mils)*** |
| **2** | **0** | **0** | **0** | **0** | **+1** | **1** |
| **3** | **-1** | **0** | **1** | **-0.5** | **+1** | **1.1** |
| **4** | **0** | **-1** | **1** | **0** | **+1** | **1** |
| **5** | **0** | **-1** | **1** | **0** | **+1** | **1** |
| **6** | **0** | **0** | **0** | **-0.5** | **+1** | **1.1** |
| **7** | **-1** | **-1** | **1.4** | **-1** | **0** | **1** |
| **8** | **0** | **-1** | **1** | **0** | **0** | **0** |
| **9** | **-1** | **-1** | **1.4** | **-0.5** | **+0.5** | **0.7** |
| **10** | **-0.5** | **+0.5** | **0.7** | **0** | **+1** | **1** |
| **11** | **-1** | **0** | **1** | **-1** | **+1** | **1.4** |
| **12** | **0** | **+1** | **1** | **-1** | **+1** | **1.4** |
| **13** | **+1** | **+1** | **1.4** | **0** | **+0.5** | **0.5** |
| **14** | **+1** | **0** | **1** | **0** | **-0.5** | **0.5** |
| **15** | **+0.5** | **0** | **0.5** | **-0.5** | **0** | **0.5** |
| **16** | **+0.5** | **-1** | **1.1** | **-0.5** | **0** | **0.5** |
| **17** | **0** | **-0.5** | **0.5** | **0** | **0** | **0** |
| **18** | **+0.5** | **0** | **0.5** | **0** | **0** | **0** |
| **19** | **0** | **0** | **0** | **0** | **0** | **0** |
| **20** | **0** | **-2** | **2** | **0** | **0** | **0** |
| **21** | **0** | **-2** | **2** | **-0.5** | **0** | **0.5** |
| **22** | **0** | **0** | **0** | **+1.5** | **0** | **1.5** |
| **23** | **0** | **0** | **0** | **0** | **+0.5** | **0.5** |
| **24** | **+1** | **0** | **1** | **0** | **+1** | **1** |
| **25** | **0** | **+1** | **1** | **0** | **+1** | **1** |
| **26** | **0** | **+1** | **1** | **+0.5** | **+1** | **1.1** |

**Table 1 – SN1 layer registration x-ray measurements**

Another observation based on the layer shifting measurement results is that these reference boards may be much better than can be expected from a typical fabrication process. That finding implies that this batch of test boards was manufactured with much better than average characteristics; thus our measurements may not represent typical performance. Previously the board manufacturer has stated that the layer shift specification was +/- 5 mils. More recently, the manufacturer has responded that a specification of +/- 3 mils might be realizable. It seems apparent that the measured channel performance would degrade significantly for a board with greater amounts of layer shifting. Therefore, it is best to employ a conservative approach and de-rate the performance of the channels slightly in anticipation of greater amounts of layer shifting. Moreover, it will likely be unsafe to rely only on measurements to predict electrical performance accurately. Instead, results from advanced simulation models that take into account the layer shifting will be required to obtain the nominal and statistical variation of the via performance.

## Channel Skew

Unmitigated fiber weave skew (FWS) can easily cause $56G_{NRZ}$ link failures. [10] explains the rationale to minimize total channel skew below 0.2 unit intervals (UI). At 56 Gb/s, the UI is 18 psec and so the channel skew budget is only 3.6 psec. FWS can only take a portion of this skew limit since there are other skew contributors.

Rotated structures described in [10] were demonstrated to have less than 185 fsec of total (1-sigma) skew, implying that both probe-launch and FWS were very low, although it was not possible to separate skew contributions from each source at these low skew levels. Therefore we repeated those FWS test structures in this present set of test boards. These structures are depicted in Figure 16-a. Since FWS is a statistical phenomenon, a reasonable number of structures are required to achieve statistically significant skew values. To that end, we chose to stagger sets of three differential striplines and repeat this pattern. For these most recent test boards, with limited board area allocated for FWS testing, we constrained the structures to seven repeated sets, or 21 total stripline pairs for both $0^o$ and $10^o$ artwork rotation.

Since we want to measure FWS of differential striplines, a ground via was placed between the complementary signal vias to limit the coupling between signal vias, which in turn could affect the FWS measurement. Also, we utilized G-S-G-G-S-G microwave probes to test the skew structures – one probe per side to perform the 4-port measurements. Four additional ground vias were also added, i.e., five in total, to the ground probe pad locations.

(a)

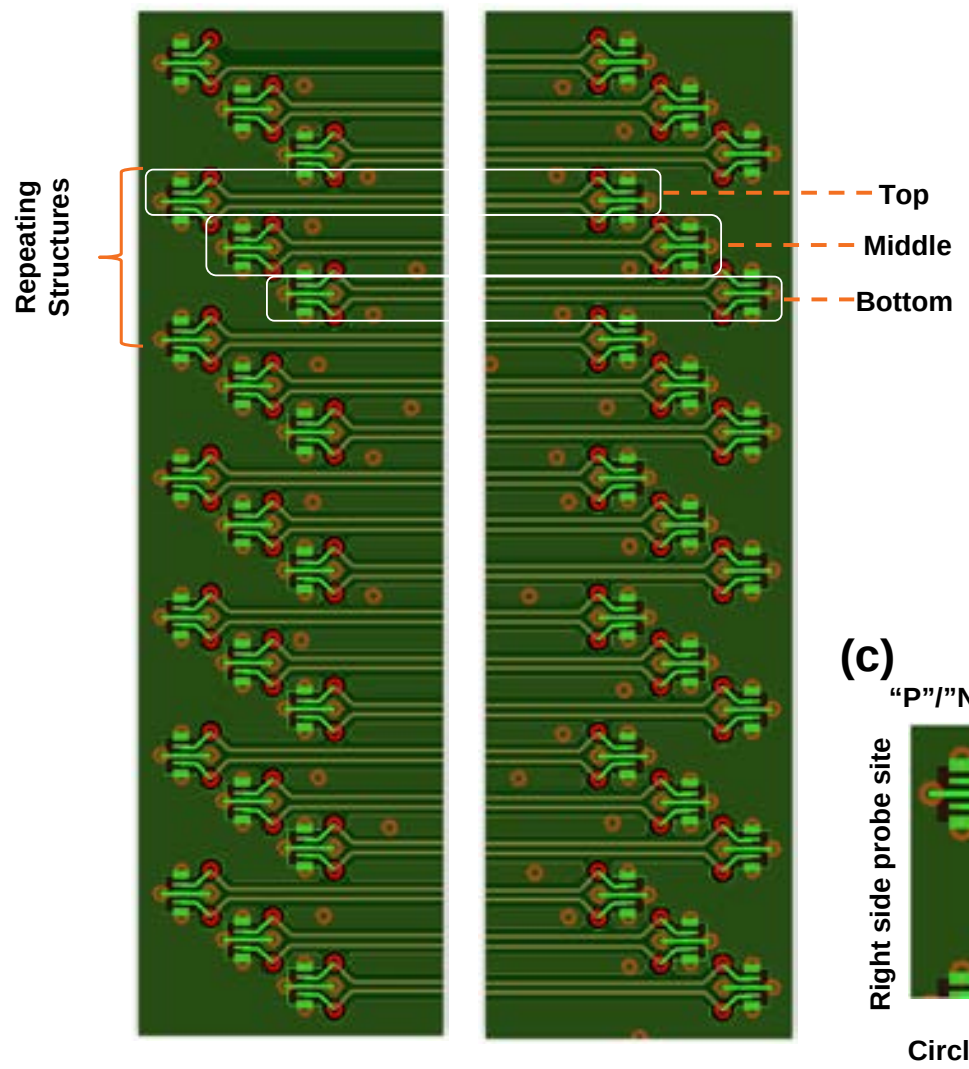

(b)

***Ground via count "close" to signal via***

| | Left Side | Right Side |
|---|---|---|
| Top P | 2 | 2 |
| Top N | 4 | 2 |
| Middle P | 2 | 4 |
| Middle N | 4 | 2 |
| Bottom P | 2 | 4 |
| Bottom N | 2 | 2 |

(c)

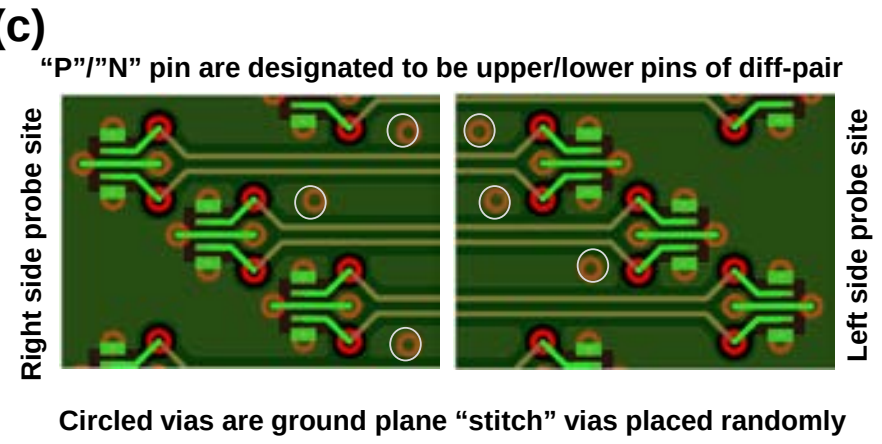

**Figure 16 - Fiber weave skew test patterns [45086]**

As with all of our previous test boards, we also added ground vias across each board to stitch together ground planes and minimize ground plane resonances. These ground vias were randomly spaced from one another so that periodic structures were avoided to minimize resonances. Many of these randomly placed vias did come in close proximity to the FWS stripline signal vias. Specifically, in Figure 16-c, we have circled these ground stitch vias. Momentarily ignoring these stitching vias, the table in Figure 16-b presents a count of the number of vias close to the signal vias for the top, middle, and bottom P/N vias in the set of three "top", "middle", and "bottom" stripline structures. Note these ground via locations; further comments in this regard will be forthcoming.

The P/N skew measurements from all five boards are presented in Figure 17-a for each structure of the 21 individual test patterns for the case with $0^o$ rotation (left side) and with $10^o$ rotation (right side). The x-axes represent trace positions for each test structure – the 21 structures within a set have sequential trace position designations. Clearly the rotated striplines exhibit significantly less skew. Both rotation cases show repeated cyclic patterns. These cyclic patterns have been observed previously [10] and are suspected to be an artifact of the "mixing" of pattern repeat pitch and the pitch of the fiberglass bundles. However, the mixing did not make sense in that the "phase" of the mixing should have been random across the five structures measured on differing boards.

Noting the differences in launch structures with respect to ground via locations for top, middle, and bottom test structures, the measured skew data is re-plotted in Figure 17-b, but this time with each structure grouped into sets of the top, middle and bottom positions. Note that the skews for the un-rotated structures are now much more random, i.e., the saw-tooth patterns are no longer present. More significantly, the skew for the structures with rotation show fairly low skew variation, but have significant mean skew ranging from +/- 2 psec. These results help to confirm that much of the skew originated from the probe launch area rather than in the fiber weave. We have measured mean skew much lower than 1-sigma random skew on previous boards using these same patterns. However, the previous boards were much thinner, on the order of 35 mils in thickness, whereas our present test boards are approximately 120 mils thick. Hence it appears that the skew effects of the imbalanced ground return paths near the signal via barrels caused the high skew measurements. It is unlikely that we would have made this observation if we had used a routing layer close to the top surface of the board.

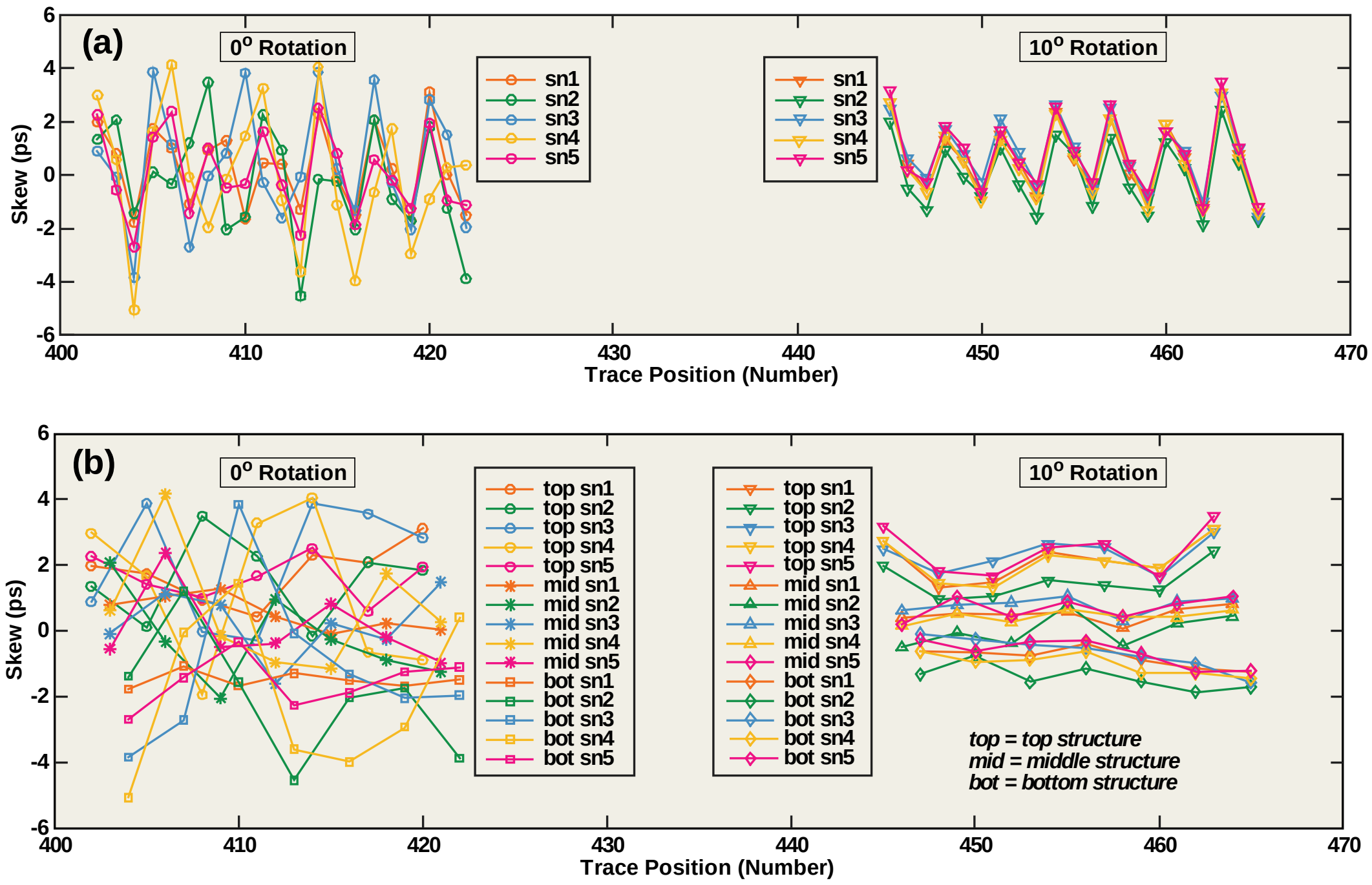

**Figure 17 - Measured skew on fiber weave skew test structures [44935]**

To investigate the high mean skew results further, we reviewed both the right and left side probe launch behavior by viewing the TDR responses. To limit FWS effects, only the rotated set of 21 structures were investigated. The TDR data by themselves all appeared very similar. For this situation the in-phase differences between P and N values within a differential structure yielded interesting results, as presented in Figure 18. The polarity of these difference waveforms follows reasonably well with the pattern of the number of ground vias in Figure 16-b. The polarities of these waveforms appear correct when we consider that the added ground vias tend to lower the inductance of the probe launch. Using Figure 18-a as an example, the P-side, having two ground return vias, is more inductive than the N-side, having four ground return vias. Since higher inductance creates a more positive TDR response (from a positive-going step) then the difference waveform should be positive.

Figure 18-b and Figure 18-e have a balanced number of vias, yet some residual skew is still in evidence. It appears that there is considerable variability across top, bottom, and middle groupings, since some of these structures have the randomly placed ground “stitching” vias nearby, while others do not have these nearby vias. Rather than plotting difference waveforms from one board as grouped into top, middle, and bottom classifications, we examined individual structures across multiple boards; these results are presented in Figure 19. For this case structures 4, 5, and 6 of the 21 structures within a set were selected for review. These are still top, middle, and bottom structures, respectively; note that the same patterns as previously described are present, but without the effects of the randomly placed stitching vias. Moreover, these results should provide some insight as to the board-to-board variability of the via launch.

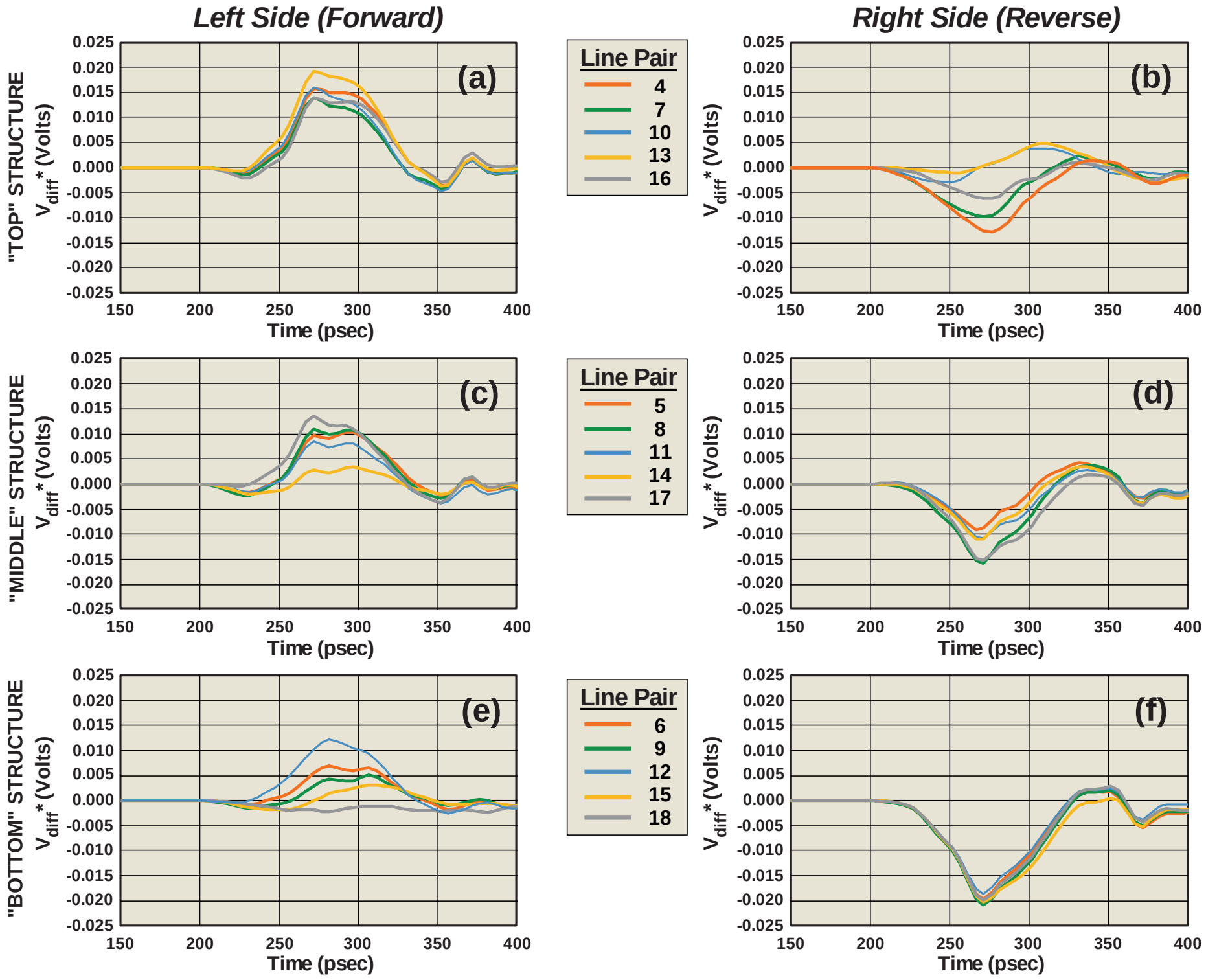

**$V_{diff}$ is the difference in voltage between P and N sides, both stimulated with 1 volt pulse*

**Figure 18 - Differences in P/N TDRs (both positive going stimuli) for FWS test structures in top, middle, and bottom structure groupings for SN1 [45087]**

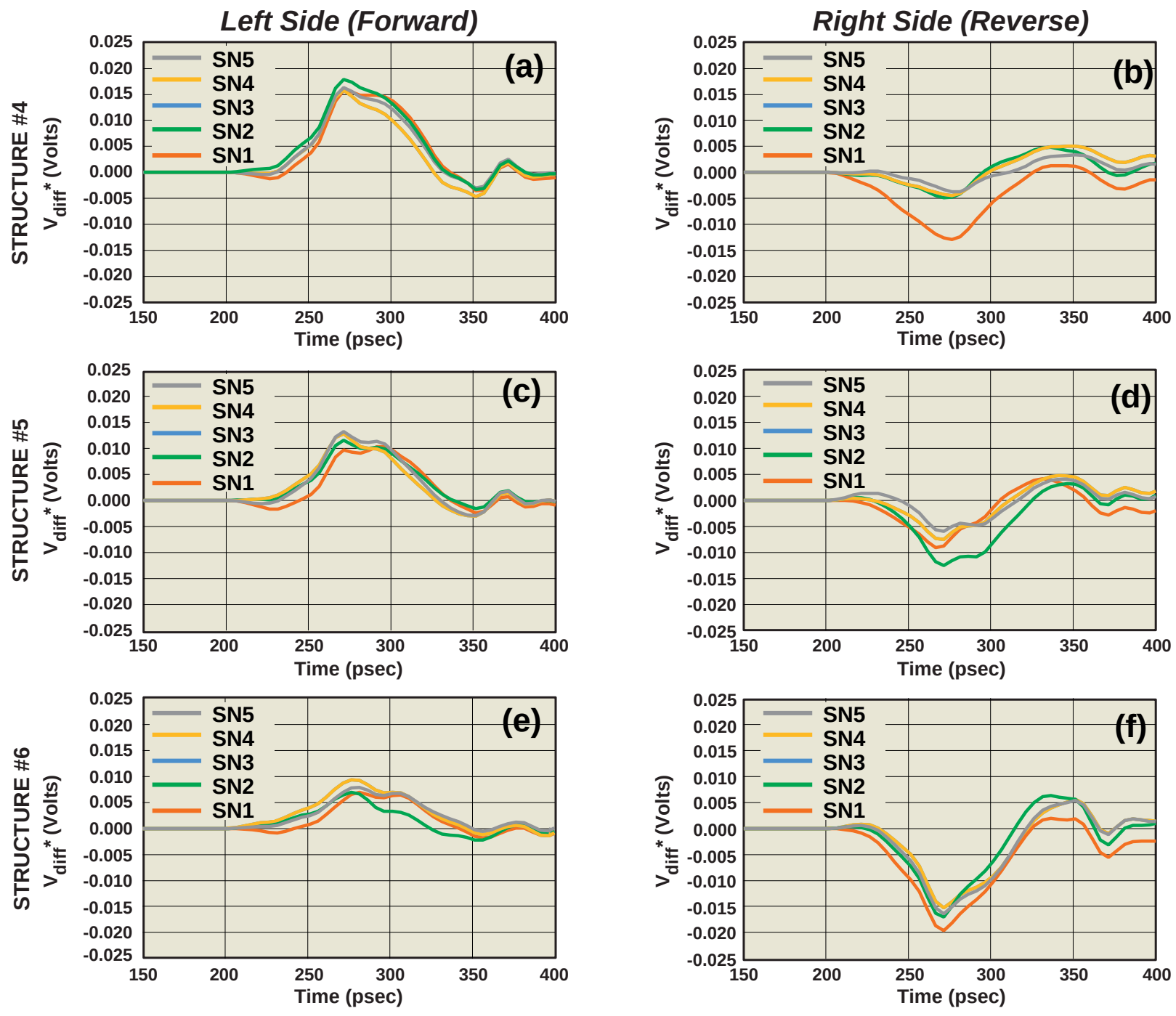

*$V_{diff}$ is the difference in voltage between P and N sides, both stimulated with 1 volt pulse*

**Figure 19 - Differences in P/N TDRs (both positive) for FWS structures 4, 5, and 6 for all five boards [45088]**

Unfortunately, the large amount of pin-field skew as well as the effects from the stitching vias reduces the ability to isolate and extract the FWS. Due to the effect of the randomly placed stitching vias each of the 21 test sites is unique and we have five measurements of each structure – one from each of the five boards. It is apparent that the standard deviation (1-sigma) skew was much lower with rotated striplines but it was not possible to differentiate between skew contributions from the pin-fields and the FWS effects of the striplines. Without rotation the skew has a 1-sigma range across the 21 structures from 0.64 to 2.33 psec. Assuming that the typical skew was the average of this range, then the 3-sigma skew was 4.5 psec, which greatly exceeds the entire skew budget of 3.9 psec. With rotation, the 1-sigma skew decreases to 0.19 to 0.49 psec. Assuming the typical skew to be an average of this range, then the 3-sigma skew is roughly 1.0 psec, leaving 2.9 psec in the skew budget for other effects. System yield goals may require a larger FWS budget (> 3-sigma) leaving less than 2.9 psec for other skew contributors. To better isolate FWS, the L3 blind via should have been used and structures should have been spaced further apart. With the limited FWS data that has been collected to date, it appears certain that artwork rotation is essential for $56G_{NRZ}$ applications.

## Summary

We identified three stripline layers within a test board that have sufficient via launch (pin-field) electrical performance to suggest that $56G_{NRZ}$ is feasible for PCB technologies without resorting to advanced processing techniques beyond the use of a few blind vias and two sub-laminations. We were also able to identify several other areas of study to improve upon via performance that may enable additional stripline layers capable of supporting $56G_{NRZ}$. Test boards developed for our studies were measured to have very minimal layer shifting during manufacture; hence the measured electrical performance may be somewhat better than that expected of a typical PCB. For the layer shifting topic, we suggest that $56G_{NRZ}$ PCB performance should be obtained from a statistical model since it will not generally be feasible to collect and measure a sufficiently large set of sample boards.

It may be difficult to meet the 3.6 psec channel skew requirements for $56G_{NRZ}$ applications. Our measurements show 2 psec of combined skew just in BGA egress and ingress routing. This same order of skew, due to pin-field ground return asymmetry, was also observed. Both of these skew contributors would benefit from further study to develop good design rules and to help quantize both design-dependent and random skew. Using rotated artwork to reduce FWS certainly appears to be a necessity to allow the channel skew budget to be met.

We did not review whether crosstalk could increase for $56G_{NRZ}$ applications to determine whether all three channel figures of merit, i.e., loss, skew, and crosstalk, could be achieved. If each of these three figures of merit cannot be resolved, then evolutionary PCB technologies, such as limited mechanically drilled blind vias, shallow laser-blind vias, and moderate (12:1) via aspect ratios will have to be replaced with better performing, perhaps revolutionary PCB technologies.